\documentclass[review,authoryear,3p,times,10pt]{elsarticle}

\usepackage{multirow,setspace,times,amssymb,amsmath,graphicx,color,rotating,subfigure,url}
\usepackage{lineno,color}
\usepackage{natbib}
\usepackage{booktabs}%
\usepackage{longtable}%
\usepackage{rotating}
\usepackage{lscape}
\usepackage{threeparttable}
\graphicspath{{Figures/}}
\usepackage[table]{xcolor}
\usepackage{tabularx}
\usepackage{graphicx} 
\usepackage{epstopdf}
\usepackage{mathrsfs}
\usepackage{makecell}
\usepackage[bookmarks=true,colorlinks,linkcolor=blue,anchorcolor=blue,citecolor=blue,unicode]{hyperref}
\usepackage{bookmark}

\usepackage{todonotes}
\usepackage{ragged2e}
\usepackage[font=small]{caption}

\hypersetup{CJKbookmarks=true}%
\journal{Journal of Economic Behavior \& Organization}

\begin{document}

\begin{frontmatter}
\title{The impact of the Russia-Ukraine conflict on the extreme risk spillovers between agricultural futures and spots}

\author[SB,RCE,DM]{Wei-Xing Zhou}
\ead{wxzhou@ecust.edu.cn}
\author[SB]{Yun-Shi Dai}
\author[SBS]{Kiet Tuan Duong}
\author[WHUT1,WHUT2]{Peng-Fei Dai\corref{CorAuth}}
\ead{pfdai@ecust.edu.cn}
\cortext[CorAuth]{Corresponding author.} 

\address[SB]{School of Business, East China University of Science and Technology, Shanghai 200237, China}
\address[RCE]{Research Center for Econophysics, East China University of Science and Technology, Shanghai 200237, China}
\address[DM]{School of Mathematics, East China University of Science and Technology, Shanghai 200237, China}
\address[SBS]{School for Business and Society, University of York, York YO10 5ZF, UK}
\address[WHUT1]{School of Management, Wuhan University of Technology, Wuhan 430070, China}
\address[WHUT2]{Research Institute of Digital Governance and Management Decision Innovation, Wuhan University of Technology, Wuhan 430070, China}

\begin{abstract}
The ongoing Russia-Ukraine conflict between two major agricultural powers has posed significant threats and challenges to the global food system and world food security. Focusing on the impact of the conflict on the global agricultural market, we propose a new analytical framework for tail dependence, and combine the Copula-CoVaR method with the ARMA-GARCH-skewed Student-t model to examine the tail dependence structure and extreme risk spillover between agricultural futures and spots over the pre- and post-outbreak periods. Our results indicate that the tail dependence structures in the futures-spot markets of soybean, maize, wheat, and rice have all reacted to the Russia-Ukraine conflict. Furthermore, the outbreak of the conflict has intensified risks of the four agricultural markets in varying degrees, with the wheat market being affected the most. Additionally, all the agricultural futures markets exhibit significant downside and upside risk spillovers to their corresponding spot markets before and after the outbreak of the conflict, whereas the strengths of these extreme risk spillover effects demonstrate significant asymmetries at the directional (downside versus upside) and temporal (pre-outbreak versus post-outbreak) levels. 
\end{abstract}


\end{frontmatter}


\section{Introduction}

Food is a basic human need, and food security is an essential issue concerning human survival and development. However, the current situation of global food security is complex and severe, with many countries and regions facing numerous risks and challenges. According to {\textit{The Global Report on Food Crises 2023}}\footnote{\url{https://www.fsinplatform.org/global-report-food-crises-2023}} jointly published by several international organizations, including the Food and Agriculture Organization of the United Nations (FAO) and the World Food Programme (WFP), approximately 258 million people in 58 countries and regions worldwide suffered from acute food insecurity in 2022. This is a noticeable increase from 193 million people in 53 countries and regions in 2021, reaching a historic high. The report also emphasizes that geopolitical conflicts, weather extremes, and economic shocks are increasingly intertwined with the lingering socioeconomic impacts of the COVID-19 pandemic, intensifying the current global food crisis. In particular, the Russia-Ukraine conflict, which broke out on February 24, 2022, has further revealed the inherent problems of interrelatedness and interdependence between global and local food systems \citep{Sun-Scherer-Zhang-Behrens-2022-NatFood}, exacerbated the vulnerability of the current world food system \citep{Behnassi-ElHaiba-2022-NatHumBehav, ReinaUsuga-ParraLopez-deHaroGimenez-CarmonaTorres-2023-LandUsePol}, and sparked grave concerns about global food security.

Russia and Ukraine, as both parties to the conflict, are not only major producers, but also important exporters of food, energy, and chemicals globally. Regarding grain production, according to the statistics released by the United States Department of Agriculture (USDA)\footnote{\url{https://apps.fas.usda.gov/psdonline/app/index.html\#/app/downloads}}, the wheat harvest of Ukraine and Russia in the 2021/22 season reached 33.01 million tons and 75.16 million tons, respectively, accounting for 4.24\% and 9.65\% of the total global wheat output. Their maize yields were 42.13 million tons and 15.23 million tons, representing 3.46\% and 1.25\% of the total global maize output, respectively. Furthermore, their barley yields were 9.92 million tons and 17.51 million tons, making up 6.80\% and 11.99\% of the total global barley output, respectively. With respect to grain trade, USDA data shows that Russia and Ukraine, as the world's largest and fifth-largest wheat exporters, exported 33 million tons and 18.84 million tons of wheat, respectively, in the 2021/22 season, accounting for 16.07\% and 9.18\% of the total global wheat exports. Moreover, Ukraine ranks third and Russia fourth in terms of barley export volumes, with Ukraine exporting 5.8 million tons and Russia exporting 4.5 million tons, making up 16.78\% and 13.02\% of the total global barley exports, respectively. Additionally, Ukraine and Russia also hold significant positions as major maize exporters worldwide. Ukraine exported 26.98 million tons of maize while Russia exported 4 million tons, representing 13.94\% and 2.07\% of the total global maize exports, respectively. Considering the indispensable roles played by Russia and Ukraine in the world food supply, the outbreak of the Russia-Ukraine conflict has undoubtedly exerted a substantial impact and posed a non-negligible risk to the global food system \citep{Li-Song-2022-Foods, Rawtani-Gupta-Khatri-Rao-Hussain-2022-SciTotalEnviron, Shumilova-Tockner-Sukhodolov-Khilchevskyi-DeMeester-Stepanenko-Trokhymenko-HernandezAgueero-Gleick-2023-NatSustain}.

As a conflict between two major agricultural powers, the Russia-Ukraine conflict has revealed inherent defects in global agricultural production and governance systems \citep{Yang-Liu-2022-Land}, and given rise to numerous serious consequences for the global food market. Following the COVID-19 pandemic, the conflict has further disrupted the global food supply chain, exerting tremendous strain on the world food supply \citep{Jagtap-Trollman-Trollman-GarciaGarcia-ParraLopez-Duong-Martindale-Munekata-Lorenzo-Hdaifeh-Hassoun-Salonitis-AfyShararah-2022-Foods, Qin-Su-Umar-Lobont-Manta-2023-EconAnalPolicy}. In addition, the adverse repercussions of this conflict on upstream energy and fertilizer markets are likely to be transmitted to the food market, which may result in more acute food insecurity in certain countries and regions \citep{Feng-Jia-Lin-2023-ChinaAgricEconRev}.
Moreover, since the outbreak of the Russia-Ukraine conflict, food trade protectionism has continued to heat up, with many countries implementing export restrictions. The resulting tension between food supply and demand has fueled concerns about global food shortages, which, in turn, have intensified global capital speculation in the food market. Under the superimposed influence of the Russia-Ukraine conflict and extreme weather events, food export restrictions and excessive futures speculation have triggered sharp increases and subsequent dramatic fluctuations in international food prices, further exacerbating risks in the global food market \citep{Neik-Siddique-Mayes-Edwards-Batley-Mabhaudhi-Song-Massawe-2023-FrontSustainFoodSyst}.

The futures and spot markets are inextricably interconnected and mutually complementary. As a crucial mechanism for price discovery, the agricultural futures market reflects market expectations regarding future supply and demand dynamics for agricultural commodities. Accordingly, agricultural futures prices play an essential role in agricultural spot pricing, and can be regarded as powerful predictors and reliable benchmarks for subsequent spot prices \citep{Noussair-Tucker-Xu-2016-JEconBehavOrgan, Arzandeh-Frank-2019-AmJAgrEcon}. However, influenced by investor sentiment, trading regulations, and other factors, futures prices may not always accurately reflect the real conditions in the spot market, and thus are not always the true prices ``discovered’’ by the futures market, especially in cases involving excessive speculation \citep{Kawaller-Koch-Koch-1987-JFinanc, He-LiuChen-Meng-Xiong-Zhang-2020-QuantFinanc}. Furthermore, movements in futures prices tend to generate market expectations for spot price fluctuations. For instance, an upward trend in futures prices often fosters the prevailing market expectations of higher spot prices, thereby exerting upward pressure on actual spot prices. In recent years, driven by various factors such as global liquidity surplus, high inflation, and geopolitical conflicts, the prices of grain futures have generally remained up, attracting substantial international capital into agricultural futures markets. The consequent excessive speculation magnifies the effect of price fluctuations \citep{Zhong-Darrat-Otero-2004-JBankFinanc, Bonnier-2021-JIntMoneyFinan, Hirota-Huber-Stockl-Sunder-2022-JEconBehavOrgan}, and the excess volatility in agricultural futures markets is expected to spill over into agricultural spot markets due to their close linkage \citep{Huynh-Burggraf-Nasir-2020-ResourPolicy, DeJong-Sonnemans-Tuinstra-2022-JEconBehavOrgan}, which may lead to limited price-discovery and risk-aversion functions of agricultural futures. Therefore, in the context of the Russia-Ukraine conflict, it is imperative to clarify the dependence structure within the international food market, assess the risks of the agricultural futures and spot markets, and investigate the risk spillovers between these markets for stabilizing the global food market, optimizing the global food governance system, and maintaining global food security.

This study focuses on the impact of the Russia-Ukraine conflict on tail dependence and risk spillovers within the international food market. Selecting soybean, maize, wheat, and rice as our research objects, we utilize the Copula-CoVaR method to examine the tail dependence structures and extreme risk spillovers between food futures and spot markets before and after the outbreak of the conflict, respectively. Given the common characteristics observed in financial data, such as leptokurtosis, autocorrelation, and heteroscedasticity, we first adopt the ARMA-GARCH-skewed Student-t model to fit the marginal distributions of agricultural returns, and then propose a new tail dependence analysis framework to describe the tail dependence structure more flexibly and accurately. Specifically, we incorporate various single copula models and self-constructed mixed copula models to fully consider possible symmetric and asymmetric tail dependencies, and determine the optimal copula model for each pair according to the Akaike information criterion (AIC). On the basis of model estimation, we quantify extreme risk measures for soybean, maize, wheat, and rice spot markets over both pre- and post-outbreak periods, including downside and upside Value-at-Risk ($VaR$), Conditional Value-at-Risk ($CoVaR$), and delta Conditional Value-at-Risk ($\Delta CoVaR$). By combining these risk measures with the actual situation, we further provide reasonable explanations for their dramatic changes, offering a deeper and more intuitive understanding of the impact of the Russia-Ukraine conflict on the global food market. In addition, the Kolmogorov-Smirnov (K-S) test is applied to evaluate the significance of downside and upside risk spillover effects before and after the outbreak of the conflict, as well as the possible asymmetries of these effects at the directional (downside versus upside) and temporal (pre-outbreak versus post-outbreak) levels.

The main contributions of this study can be summarized in the following three aspects. First, this paper further enriches and expands the literature on the Russia-Ukraine conflict. Currently, the conflict is ongoing and its impact is escalating. Scholars have begun to explore the serious consequences of the Russia-Ukraine conflict from different perspectives, but few studies have quantitatively analyzed the influence of the conflict on the intrinsic connection among different food submarkets with emphasis on the interior of the global food market. Hence, our paper can serve as a supplement in this respect. Second, to the best of our knowledge, this paper is the first to examine the dependence structure and risk transmission among the agricultural futures-spot markets in the context of the Russia-Ukraine conflict. Identifying the tail dependence and quantifying the extreme risk spillover between agricultural futures and spots are conducive to expounding the specific manifestations of the impact of the conflict on the global food market, and clarifying the extreme and systemic risks of the global food system. Finally, the part about the tail dependence structure and extreme risk spillover in our paper can be taken as a reference for regulating agricultural futures markets, addressing excess volatility in food prices, and stabilizing international and domestic food markets. It also provides valuable insights for producers, consumers, and investors on using agricultural futures to transfer price risks and optimize asset allocation.

The remainder of this study is organized as follows. Section~\ref{S1:LitRev} reviews the relevant literature. Section~\ref{S1:Methodology} introduces the methodology of model construction and parameter estimation. Section~\ref{S1:Data} provides the data source and reports the statistical description. Section~\ref{S1:EmpAnal} empirically analyzes the results of the tail dependence structure and extreme risk spillover among the agricultural futures-spot markets before and after the outbreak of the Russia-Ukraine conflict. Section~\ref{S1:Conclude} concludes the findings and proposes some implications.

\section{Literature review}
\label{S1:LitRev}

On February 24, 2022, the Russia-Ukraine conflict officially broke out on a large scale. Since then, the United States, Europe, and many other countries have imposed multiple rounds of sanctions on Russia, which have generally been characterized by a broad scope and strong force \citep{Huynh-Hoang-Ongena-2022-CEPR, Gaur-Settles-Vaatanen-2023-JManageStud}. These sanctions cover various areas, such as finance, technology, energy, and food, targeting individuals, enterprises, and governments. In response, Russia has implemented a series of anti-sanction measures against the Western countries. The direct costs of the Russia-Ukraine conflict, combined with the escalating game between sanctions and anti-sanctions, have not only damaged the economic and livelihood situations of Russia and Ukraine, but also exerted serious impacts on global financial, energy, food, and other sectors.

The Russia-Ukraine conflict and subsequent sanctions have exacerbated the instability of the international financial market, altered the global financial landscape, and dealt a severe blow to the global financial order. Against this backdrop, scholars have extensively examined the implications of the conflict for the world economy. \cite{Qureshi-Rizwan-Ahmad-Ashraf-2022-FinancResLett} explore the effect of the Russia-Ukraine conflict on the systemic risks of multiple countries based on a developed database of news events, revealing an intensification of systemic vulnerability in the world financial system. \cite{Liadze-Macchiarelli-MortimerLee-Juanino-2023-WorldEcon} assess the huge costs of the Russia-Ukraine conflict and further predict the severe consequences of the conflict on the world economy, including shrinking Gross Domestic Product and increasing inflation. \cite{Tajaddini-Gholipour-2023-IntRevFinanc} empirically analyze the impact of the conflict on international trade and stock markets, and find that countries with higher trade dependence on Ukraine and Russia suffer greater stock market decline, and the strength of this relationship is influenced by the extent of national trade openness. \cite{Balli-Billah-Chowdhury-2022-TourEcon} and \cite{Pandey-Kumar-2023-CurrIssuesTour} focus on the impacts of the Russia-Ukraine conflict on the global hospitality and tourism stock sector markets, respectively, portraying a sharp increase in return connectivity and the different reactions exhibited by companies in different regions. \cite{Yousaf-Patel-Yarovaya-2022-JBehavExpFinanc} evaluate the influence of the onset of the Russia-Ukraine conflict on various equity markets at the overall, country, and regional levels. \cite{Ahmed-Hasan-Kamal-2022-EurFinancManag} capture noticeable negative abnormal returns in European equity markets following the onset of the conflict, and the extent of the negative reactions varies by country, industry, and firm size. \cite{Martins-Correia-Gouveia-2023-JMultinatlFinancManag} explore the short-term effect of the Russia-Ukraine conflict on the stocks of European listed banks. \cite{Korovkin-Makarin-2023-AmEconRev} examine trade transactions in Ukraine around the 2014 Russia-Ukraine conflict, and find that the conflict can cause economic damage even far from the fighting areas through social impacts.

From a global perspective, Russia and Ukraine are important exporters of several bulk commodities, including energy, metals, and agricultural products. As a result, the outbreak of the Russia-Ukraine conflict and the sanctions imposed on Russia have aggravated supply risks and triggered grave concerns about shortages, leading to rapid rises and subsequent dramatic fluctuations in the prices of relevant commodities. Considering the far-reaching influence of the Russia-Ukraine conflict, how the international commodity market reacts to the onset of the conflict has become a hot topic that has aroused wide attention and discussion. \cite{Adekoya-Oliyide-Yaya-AlFaryan-2022-ResourPolicy,Adekoya-Asl-Oliyide-Izadi-2023-ResourPolicy} explore the linkages between oil and important financial assets such as stocks and bonds, and conclude that oil shows strong spillovers on all the selected financial assets during the Russia-Ukraine conflict. \cite{Ha-2023-EnvironSciPollutRes} reveals the interconnections between the crude oil, gold, and stock markets before and after the conflict, thus determining the sources of fluctuations in the oil market. \cite{Sokhanvar-Lee-2023-EmpirEcon} empirically analyze the impact of energy price movements on exchange rates during the conflict. \cite{Umar-Riaz-Yousaf-2022-ResourPolicy} examine the influence of the conflict on the renewable energy, traditional energy, and metal markets, and suggest that the abnormal returns of the renewable energy sector increased markedly following the onset of the Russia-Ukraine conflict. \cite{Jahanshahi-Uzun-Kacar-Yao-Alassafi-2022-Mathematics} combine deep learning algorithms and machine learning to develop a novel program to forecast oil prices in the context of the Russia-Ukraine conflict and the COVID-19 pandemic. \cite{Steffen-Patt-2022-EnergyResSocSci} consider the Russia-Ukraine conflict as a possible milestone for European energy policies and provide evidence pertaining to how the conflict gradually changes public opinion. \cite{Sokhanvar-Bouri-2023-BorsaIstanbRev} focus on the price increases in commodity markets caused by the Russia-Ukraine conflict, and empirically analyze the impact of commodity price rises on the selected exchange rate markets. \cite{Cui-Maghyereh-2023-IntRevFinancAnal} explore the risk connectivity among international commodity futures and oil futures after the onset of the Russia-Ukraine conflict, finding time-varying positive correlations between commodity and oil futures, which have strengthened with the outbreak of the conflict. 
\cite{Zhang-Yang-Hu-Jiao-Wang-2023-EnergyEcon} unveil the impact and transmission channels of the Russia-Ukraine conflict on oil prices.

Recognizing the Russia-Ukraine conflict as a confrontation between two major agricultural powers, several scholars have emphasized its impact on the global food market and food security. \cite{BenHassen-ElBilali-2022-Foods} elaborate on the direct and indirect influences of the conflict on global food security, highlighting the need for a transition from the current inefficient, fragile, and rigid global food system to an equitable, healthy, and sustainable food system, which is in line with the viewpoint of \cite{Poertner-Lambrecht-Springmann-Bodirsky-Gaupp-Freund-LotzeCampen-Gabrysch-2022-OneEarth}. \cite{Just-Echaust-2022-EconLett} and \cite{Fang-Shao-2022-FinancResLett} explore risk spillovers and volatility risks in global agricultural markets during the ongoing Russia-Ukraine conflict, and their findings demonstrate a substantial increase in uncertainty and volatility in global food markets. \cite{Feng-Jia-Lin-2023-ChinaAgricEconRev} employ a structural general equilibrium trade framework to quantify the possible negative effects of the conflict on food production, food prices, food trade, and food security. \cite{Zhou-Lu-Xu-Yan-Khu-Yang-Zhao-2023-ResourConservRecycl} develop an underload cascading failure model and a new simulation approach to identify the magnitude and process of the impact of the Russia-Ukraine conflict on global food security. Given the uncertain duration of the conflict, \cite{Lin-Li-Jia-Feng-Huang-Huang-Fan-Ciais-Song-2023-GlobFoodSecur-AgricPolicy} simulate the impact of the Russia-Ukraine conflict on the global wheat market under mild, moderate, and severe scenarios, respectively, and further examine the global food insecurity triggered by the conflict. \cite{Arndt-Diao-Dorosh-Pauw-Thurlow-2023-GlobFoodSecur-AgricPolicy} quantitatively analyze the impact of the conflict and its resultant commodity price increases on poverty, hunger, agri-food systems, and food security in developing countries.

Correlation analysis and risk contagion have always been core issues in risk management practices, and have inspired considerable research using various techniques and tools \citep{Bae-Karolyi-Stulz-2003-RevFinancStud, Aloui-BenAissa-Nguyen-2011-JBankFinanc, Adrian-Brunnermeier-2016-AmEconRev, Ji-Liu-Zhao-Fan-2020-IntRevFinancAnal}. In recent years, with the continuous development of commodity financialization \citep{Bruno-Buyuksahin-Robe-2017-AmJAgrEcon, Awasthi-Ahmad-Rahman-Phani-2020-ResourPolicy}, the targeted research subjects have gradually expanded from traditional financial markets to emerging commodity markets. \cite{Gozgor-Lau-Bilgin-2016-JIntFinancMarkInstMoney} and \cite{Kumar-Tiwari-Raheem-Hille-2021-ResourPolicy} examine the volatility transmissions and dynamic correlations between the crude oil and agricultural commodity markets, respectively. \cite{Ji-Bouri-Roubaud-Shahzad-2018-EnergyEcon} adopt a time-varying copula model with a switching dependence to investigate tail dependencies and risk spillovers among the energy and agricultural commodity markets. \cite{Bonato-2019-JIntFinancMarkInstMoney} explores dynamic price correlations and volatility spillovers within commodity markets, including soft and grain commodities, and between oil and agricultural commodities. \cite{Dai-Dai-Zhou-2023-JIntFinancMarkInstMoney} utilize the daily prices of major agricultural products from 2000 to 2022 to reveal asymmetric dependencies and risk spillovers across diverse agricultural markets. \cite{Zhang-Yang-Li-Hao-2023-JFuturesMark} elucidate simultaneous and non-simultaneous spillovers of lower- and higher-moment risks in commodity markets, highlighting the characteristics of different commodity categories.

According to our review and synthesis of relevant literature, existing studies regarding the Russia-Ukraine conflict and sanctions tend to focus mainly on their knock-on effects on food prices, food trade, and other external manifestations, but ignore the impact of the conflict on the dependence structure and risk spillover across different food submarkets within the global food market. To make a conscious effort to fill this gap, we empirically examine the tail dependence structures and extreme risk spillovers between agricultural futures and spots before and after the outbreak of the conflict. With an emphasis on the interior of the global food market, this study is expected to clarify and elucidate the specific characterizations about the impact of the Russia-Ukraine conflict on the internal structure as well as the extreme and systemic risks of the global food market.

\section{Methodology}
\label{S1:Methodology}

\subsection{Marginal distribution modelling}

Let $r_{1}$ and $r_{2}$ refer to the futures and spot return series for each agricultural commodity, respectively. Given that the returns of financial assets tend to exhibit the characteristics of leptokurtosis and skewness, as well as autocorrelation and heteroscedasticity, the ARMA$(m, n)$-GARCH$(p, q)$-skewed Student-t model is employed in our paper to specify the marginal distributions of agricultural returns, the expression of which is given by
\begin{subequations}
  \begin{equation}
    r_{i,t} = \varphi_{0} + \sum\limits_{j=1}^{m} \varphi_{j}r_{i,t-j} + \varepsilon_{i,t} + \sum\limits_{j=1}^{n} \gamma_{j}\varepsilon_{i,t-j},\ i=1,2,\ t=1,\dots,T
    \label{Eq:Marginal_distribution_return},
  \end{equation}
  \begin{equation}
    \varepsilon_{i,t} = \sigma_{i,t}z_{i,t},\ z_{i,t} \sim i.i.d.skst_{v_{i}},
    \label{Eq:Marginal_distribution_error_term}
  \end{equation}
  \begin{equation}
    \sigma_{i,t}^{2} = \alpha_{0} + \sum\limits_{j=1}^{p} \alpha_{j}\varepsilon_{i,t-j}^{2} + \sum\limits_{j=1}^{q} \beta_{j}\sigma_{i,t-j}^{2},
    \label{Eq:Marginal_distribution_conditional_variance}
  \end{equation} 
  \label{Eq:Marginal_distribution}%
\end{subequations}
where $z_{i,t}$ denotes the standardized residual which follows the skewed Student-t distribution with $v_{i}$ degrees of freedom, and $\varepsilon_{i,t}$ and $\sigma_{i,t}^{2}$ refer to the error term and the conditional variance, respectively.

Referring to \cite{Hansen-1994-IntEconRev}, the skewed Student-t distribution takes nonzero skewness and excess kurtosis into consideration, the density function of which is expressed as
\begin{equation}
    f\left( z_{t} \mid \nu,\eta \right) = \left\{
    \begin{aligned}
    bc\left[ 1 + \frac{1}{\nu-2}\left( {\frac{bz_{t}+a}{1-\eta}} \right)^{2} \right]^{-(\nu+1)/2}, \ z_{t} < -\frac{a}{b} \\
    bc\left[ 1 + \frac{1}{\nu-2}\left( {\frac{bz_{t}+a}{1+\eta}} \right)^{2} \right]^{-(\nu+1)/2}, \ z_{t} \geq -\frac{a}{b} 
    \end{aligned}
    \right.
    \label{Eq:The_skewed_Student-t_density_distribution}
\end{equation}
where $\eta$ and $\nu$ refer to the asymmetric and degrees-of-freedom parameters, respectively, and the constant terms $a$, $b$, and $c$ are computed by 
\begin{subequations}
\begin{equation}
    a = 4\eta c\left( \frac{\nu-2}{\nu-1} \right),
    \label{Eq:The_skewed_Student-t_density_distribution_a}
\end{equation}
\begin{equation}
    b^{2} = 1+3\eta^{2}-a^{2},
    \label{Eq:The_skewed_Student-t_density_distribution_b}
\end{equation}
and
\begin{equation}
    c = \frac{\Gamma\left( \frac{\nu+1}{2} \right)} {\sqrt{\pi(\nu-2)}\Gamma\left( \frac{\nu}{2} \right)}.
    \label{Eq:The_skewed_Student-t_density_distribution_c}
\end{equation}
\label{Eq:The_skewed_Student-t_density_distribution_abc}%
\end{subequations}

\subsection{Copula modelling}

A copula can be interpreted as a multivariate cumulative distribution function with uniform marginal distribution functions over the interval [0, 1], which describes the dependence or association between multiple variables and helps to isolate the marginal or joint probabilities of variable pairs. Different copula functions can capture various dependence structures when specifying joint distributions, making it feasible to allow for diverse dependencies. Furthermore, there is no restriction on the choice of marginal distributions in copula modelling, which contributes to more accurate and realistic specifications of marginal models. Moreover, owing to its flexibility \citep{Patton-2012-JMultivarAnal}, the copula approach has significant advantages in describing nonlinear and tail dependencies, particularly in the case of extreme price movements. Therefore, the copula method has been widely utilized in econometric studies to depict the dependence between various assets and markets \citep{Patton-2006-IntEconRev, Christoffersen-Errunza-Jacobs-Langlois-2012-RevFinancStud, Bollinger-Hirsch-Hokayem-Ziliak-2019-JPolitEcon}.

Based on Sklar's theorem developed by \cite{Sklar-1959-PublInstStatistUnivParis}, each multivariate joint distribution function can be decomposed into a unique copula function and several univariate marginal distribution functions. Accordingly, the joint distribution function $F$ of the bivariate return series $\mathbf{r}_{t} = \left( r_{1,t},r_{2,t}\right )$ can be expressed as
\begin{equation}
    F\left( r_{1,t},r_{2,t};\theta \right) = C_{t} \left( F_{1}\left(r_{1,t};\theta_{1} \right),F_{2}\left( r_{2,t};\theta_{2} \right);\theta_{c} \right),\ \theta = \left(\theta_{1}^{\prime},\theta_{2}^{\prime},\theta_{c}^{\prime}\right)^{\prime}
    \label{Eq:Joint_distribution_function}
\end{equation}
where $C_{t}$ is the copula distribution function, and $F_{1}$ and $F_{2}$ refer to the marginal distribution functions of the return series $r_{1}$ and $r_{2}$, respectively.

Given the assumption that all the cumulative distribution functions (CDF) are differentiable, the joint density function $f$ of the bivariate return series $\mathbf{r}_{t} = \left( r_{1,t},r_{2,t}\right )$ can be obtained by
\begin{equation}
    f\left( r_{1,t},r_{2,t};\theta \right) = c_{t} \left( F_{1}\left(r_{1,t};\theta_{1} \right),F_{2}\left( r_{2,t};\theta_{2} \right);\theta_{c} \right) \cdot f_{1}\left(r_{1,t};\theta_{1} \right) \cdot f_{2}\left( r_{2,t};\theta_{2}\right),
    \label{Eq:Joint_density_function}
\end{equation}
where $c_ {t} $ is the copula density function, and $f_{1}$ and $f_{2}$ denote the marginal density functions of the return series $r_{1}$ and $r_{2}$, respectively.

\subsubsection{Single copula model}

To capture various dependence characteristics between agricultural futures and spot returns, we adopt different single copula models in our analytical framework, including the bivariate Normal, Student-t, Clayton, survival Clayton, Gumbel, and survival Gumbel copulas.

Normal copula describes tail independence, whose distribution function is 
\begin{equation}
    C_{N}\left( u_{1},u_{2}; \rho \right) = \int\nolimits_{-\infty}^{\phi^{-1} \left( u_{1} \right) } \int\nolimits_{-\infty}^{\phi^{-1} \left( u_{2} \right) } \frac{1}{2\pi\sqrt{1-\rho^{2}}} \exp \left\{-\frac{s_{1}^{2}+s_{2}^{2}-2\rho s_{1} s_{2}}{2 \left(1-\rho^{2} \right)} \right\} \mathrm{d}s_1\mathrm{d}s_2,\ \rho \in (-1, 1)
    \label{Eq:Normal_copula}
\end{equation}
where $\rho$ denotes the copula parameter, and $\phi^{-1}$ is the inverse of the CDF of the normal distribution. $u_{1}$ and $u_{2}$ refer to the probability integral transforms of $r_{1}$ and $r_{2}$ by their respective marginal distributions. The lower and upper tail dependence coefficients for the bivariate Normal copula, namely $\lambda_{N} ^ {\mathrm{low}}$ and $\lambda_ {N} ^ {\mathrm{up}}$, are both equal to 0.

Student-t copula captures symmetric tail dependence, whose distribution function is 
\begin{equation}
    C_{S}\left( u_{1},u_{2}; \rho, \nu \right) = \int\nolimits_{-\infty}^{T_{\nu}^{-1} \left( u_{1} \right) } \int\nolimits_{-\infty}^{T_{\nu}^{-1} \left( u_{2} \right) } \frac{1}{2\pi\sqrt{1-\rho^{2}}} \left[ 1 + \frac{s_{1}^{2}+s_{2}^{2}-2\rho s_{1}s_{2}}{\nu\left( 1-\rho^{2} \right)} \right]^{-\frac{\nu+2}{2}} \mathrm{d}s_1\mathrm{d}s_2,\ \rho \in (-1, 1)
    \label{Eq:Student_copula}
\end{equation}
where $\rho$ and $\nu$ denote the copula parameter and the degrees-of-freedom parameter, respectively, and $T_{\nu}^{-1}$ is the inverse of the CDF of the Student't distribution. The lower and upper tail dependence coefficients for the bivariate Student-t copula, that is $\lambda_{S}^{\mathrm{low}}$ and $\lambda_{S}^{\mathrm{up}}$, are both $2T_{\nu+1}\left( -\sqrt{\nu+1}\sqrt{\left( 1-\rho \right)/\left( 1+\rho \right)} \right)$, where $T_{\nu+1}$ represents the CDF of the Student't distribution with $\nu+1$ degrees of freedom.

Clayton copula is capable of describing lower tail dependence, whose distribution function is
\begin{equation}
    C_{C}\left( u_{1},u_{2}; \alpha \right) = \left( u_{1}^{-\alpha} + u_{2}^{-\alpha} - 1 \right)^{-\frac{1}{\alpha}},\ \alpha \in (0, +\infty)
    \label{Eq:Clayton_copula}
\end{equation}
where $\alpha$ is the copula parameter. The lower tail dependence coefficient for the bivariate Clayton copula $\lambda_{C}^{\mathrm{low}}$ is $2^{-1/ \alpha}$, and the upper tail dependence coefficient $\lambda_{C}^{\mathrm{up}}$ equals 0.

With a mirror image to the Clayton copula, the survival Clayton copula (180-degree rotated Clayton copula) can capture upper tail dependence, whose distribution function is
\begin{equation}
    C_{SC}\left( u_{1}, u_{2}; \alpha \right) = u_{1} + u_{2} - 1 + \left[ \left(1-u_{1}\right)^{-\alpha} + \left( 1-u_{2}\right)^{-\alpha} - 1 \right]^{-\frac{1}{\alpha}},\ \alpha \in (0, +\infty)
    \label{Eq:Survival_Clayton_copula}
\end{equation}
where $\alpha$ is the copula parameter. The lower tail dependence coefficient for the bivariate survival Clayton copula $\lambda_{SC}^{\mathrm{low}}$ equals 0, and the upper tail dependence coefficient $\lambda_{SC}^{\mathrm{up}}$ is $2^{-1/ \alpha}$.

Gumbel copula is capable of depicting upper tail dependence, whose distribution function is
\begin{equation}
    C_{G}\left( u_{1},u_{2}; \alpha \right) = \exp \left\{ -\left[ \left( -\ln u_{1} \right)^{\alpha} + \left( -\ln u_{2} \right)^{\alpha} \right]^{\frac{1}{\alpha}}  \right\},\ \alpha \in (1, +\infty)
    \label{Eq:Gumbel_copula}
\end{equation}
where $\alpha$ is the copula parameter. The lower tail dependence coefficient for the bivariate Gumbel copula $\lambda_{G}^{\mathrm{low}}$ is 0, and the upper tail dependence coefficient $\lambda_{G}^{\mathrm{up}}$ equals $2-2^{1/ \alpha}$.

 As a survival function of Gumbel copula, the survival Gumbel copula (180-degree rotated Gumbel copula) can describe lower tail dependence, whose distribution function is
\begin{equation}
    C_{SG}\left( u_{1}, u_{2}; \alpha \right) = u_{1} + u_{2} - 1 + \exp \left\{ -\left[ \left( -\ln (1-u_{1}) \right)^{\alpha} + \left( -\ln (1-u_{2}) \right)^{\alpha} \right]^{\frac{1}{\alpha}}  \right\},\ \alpha \in (1, +\infty)
    \label{Eq:Survival_Gumbel_copula}
\end{equation}
where $\alpha$ is the copula parameter. The lower tail dependence coefficient for the bivariate survival Gumbel copula $\lambda_{SG}^{\mathrm{low}}$ equals $2-2^{1/ \alpha}$, and the upper tail dependence coefficient $\lambda_{SG}^{\mathrm{up}}$ is 0.

\subsubsection{Mixed copula model}

Given the limitations of single copula models in describing asymmetric tail dependence, we further construct various mixed copula models and incorporate them into our analytical framework to assess the lower and upper tail dependence simultaneously and clarify the tail dependence structure between the international agricultural futures and spot markets in a more accurate and comprehensive way.

Referring to \cite{Nelsen-2006}, a mixed copula can be regarded as a convex combination of finite single copulas. Accordingly, the distribution function of a mixed copula consisting of $N$ single copulas is given by
\begin{equation}
    C_{M}\left( u_{1}, u_{2}; \theta_{M} \right) = \sum\limits_{i=1}^{N} \omega_{i} C_{i} \left( u_{1}, u_{2}; \theta_{c}^{i} \right),\ \theta_{M} = \left( \left( \theta_{c}^{1} \right)^{\prime},\cdots,\left( \theta_{c}^{N} \right)^{\prime},\omega_{1},\cdots,\omega_{N} \right) ^{\prime}
    \label{Eq:Mixed_copula}
\end{equation}
where $\theta_{c}^{i}$ and $\omega_{i}$ denote the copula parameter set and the weight parameter corresponding to the $i$-th single copula, respectively, and $\omega_{i}$ meets the conditions of $0 \leq \omega_{i} \leq 1$ and $\sum\nolimits_{i=1}^{N} \omega_{i}=1$.

With a mixed copula in the model, the joint distribution function $F$ and joint density function $f$ of the bivariate return series $\mathbf{r}_{t} = \left( r_{1,t},r_{2,t} \right)$ can be converted into
\begin{equation}
    F\left( r_{1,t},r_{2,t};\theta \right) = \sum\limits_{i=1}^{N} \omega_{i} C_{i,t} \left( F_{1}\left(r_{1,t};\theta_{1} \right),F_{2}\left( r_{2,t};\theta_{2} \right); \theta_{c}^{i} \right),
    \label{Eq:Joint_distribution_function_mixed_copula}
\end{equation}
and
\begin{equation}
    f\left( r_{1,t},r_{2,t};\theta \right) = f_{1}\left(r_{1,t};\theta_{1} \right) \cdot f_{2}\left( r_{2,t};\theta_{2}\right) \cdot \sum\limits_{i=1}^{N} \omega_{i} c_{i,t} \left( F_{1}\left(r_{1,t};\theta_{1} \right),F_{2}\left( r_{2,t};\theta_{2} \right);\theta_{c}^{i} \right),
    \label{Eq:Joint_density_function_mixed_copula}
\end{equation}
respectively, where $C_{i,t}$ and $c_{i,t}$ refer to the distribution and density functions for the $i$-th single copula.

The logarithmic likelihood function of Eq. (\ref{Eq:Joint_density_function_mixed_copula}) is given by
\begin{equation}
    L\left( \Theta \right) = L_{c} \left( \psi_{1} \right) + L_{1}\left( \psi_{2,1} \right) + L_{2}\left( \psi_{2,2} \right),\ \Theta = \left( \theta_{1}^{\prime},\theta_{2}^{\prime},\left( \theta_{c}^{1} \right)^{\prime},\cdots,\left( \theta_{c}^{N} \right)^{\prime},\omega_{1},\cdots,\omega_{N} \right)^{\prime}
    \label{Eq:Log_likelihood_joint_density_function}
\end{equation}
where $L_{c} \left( \psi_{1} \right)$, $L_{1}\left( \psi_{2,1} \right)$, and $L_{2}\left( \psi_{2,2} \right)$ denote the logarithms of the copula density function and marginal density functions of $r_{1}$ and $r_{2}$, respectively, which can be obtained by
\begin{subequations}
\begin{equation}
    L_{c} \left( \psi_{1} \right) = \sum\limits_{t=1}^{T} \log \left\{ \sum\limits_{i=1}^{N} \omega_{i} c_{i,t} \left( F_{1}\left(r_{1,t};\theta_{1} \right),F_{2}\left( r_{2,t};\theta_{2} \right);\theta_{c}^{i} \right) \right\},
    \label{Eq:Log_likelihood_copula_density_function}
\end{equation}
\begin{equation}
    L_{1}\left( \psi_{2,1} \right) = \sum\limits_{t=1}^{T} \log \left\{ f_{1} \left(r_{1,t};\theta_{1} \right) \right\},
    \label{Eq:Log_likelihood_marginal_density_r1_function}
\end{equation}
and
\begin{equation}
    L_{2}\left( \psi_{2,2} \right) = \sum\limits_{t=1}^{T} \log \left\{ f_{2} \left(r_{2,t};\theta_{2} \right) \right\},
    \label{Eq:Log_likelihood_marginal_density_r2_function}
\end{equation}
\label{Eq:Logarithms_subequations}%
\end{subequations}
where $\psi_{1} =\theta_M= \left( \left( \theta_{c}^{1} \right)^{\prime},\cdots,\left( \theta_{c}^{N} \right)^{\prime},\omega_{1},\cdots,\omega_{N} \right) ^{\prime}$, $\psi_{2,1} = \theta_{1}^{\prime}$, and $\psi_{2,2} = \theta_{2}^{\prime}$.

\subsection{Model estimation}

On the basis of the marginal and copula model construction, we utilize the Inference For the Margins (IFM) put forward by \cite{Joe_Xu_1996} for model estimation. The IFM approach is divided into two steps: first, estimating the parameters of the marginal models, and second, estimating the parameters of the copula models with the obtained marginal parameters.

Considering the difficulty of obtaining a uniform distribution for the standardized residuals with a specific distribution, we employ the Canonical Maximum Likelihood (CML) method for the transformation, which is also adopted by \cite{Wang-Wu-Lai-2013-JBankFinanc} and \cite{Ji-Bouri-Roubaud-Shahzad-2018-EnergyEcon}. The transformation formula is given by
\begin{equation}
    \hat{F}_{k}(x) = \frac{1}{T+1} \sum\limits_{t=1}^{T} I \left( \hat{\eta}_{k,t} \leq x \right),\ k=1,2
    \label{Eq:Empirical_marginal_cumulative_distribution_function}
\end{equation}
where $I(\cdot)$ represents the indicator function, the value of which is 1 if $\hat{\eta}_{k,t} \leq x$ and 0 otherwise. The cumulative probability of the $j$-th observation of $\hat{\eta}_{k,t}$ is calculated as
\begin{equation}
    \hat{u}_{k,j} = \hat{F}_{k} \left( \hat{\eta}_{k,j} \right),\ j=1,2,\dots,T.
    \label{Eq:Cumulative_probability_for_observation}
\end{equation}

Given the estimates of the marginal parameters, the copula parameter set $\psi_{1} = \left( \left( \theta_{c}^{1} \right)^{\prime}, \cdots, \left( \theta_{c}^{N} \right)^{\prime}, \omega_{1}, \cdots, \omega_{N} \right) ^{\prime}$ can be estimated by maximizing the logarithmic likelihood function $L_{c}(\psi_{1})$:
\begin{equation}
    \psi_{1} = \arg \mathop{ \max}_{\psi_{1}} L_{c}(\psi_{1}).
    \label{Eq:Maximize_the_log-likelihood_function}
\end{equation}

\subsection{Risk spillover measurement}

With the obtained parameter estimates of the marginal and copula models, we further quantify the extreme downside and upside risk measures for the agricultural returns, including $VaR$, $CoVaR$, and $\Delta CoVaR$.

$VaR$ quantifies the maximum loss of an asset over a specific period with a specific probability of occurrence. Accordingly, the downside and upside $VaR$s for $r_{i}$ at their respective confidence levels, namely $1-\alpha_{i}^{d}$ and $1-\alpha_{i}^{u}$, are given by
\begin{subequations}
  \begin{equation}
    \text{Pr} \left( r_{i,t} \leq VaR_{\alpha_{i}^{d},t}^{r_{i,t}} \right) = \alpha_{i}^{d}, 
    \label{Eq:Downside_VaR}
  \end{equation}
and
  \begin{equation}
    \text{Pr} \left( r_{i,t} \geq VaR_{\alpha_{i}^{u},t}^{r_{i,t}} \right) = \alpha_{i}^{u}, 
    \label{Eq:Upside_VaR}
  \end{equation}
\end{subequations}
where $i=1,2$, $t=1, \cdots, T$. The values of $\alpha_{i}^{d}$ and $\alpha_{i}^{u}$ are set to 0.05 and 0.95, respectively, which refer to the fifth and ninety-fifth quantiles of the return distribution.

Combined with the estimated results of the marginal ARMA$(m, n)$-GARCH$(p, q)$-skewed Student-t model, the $VaR$s for $r_{i}$ are expressed as
\begin{subequations}
\begin{equation}
    VaR_{\alpha_{i},t}^{r_{i,t}} = \mu_{i,t}+\sigma_{i,t}t_{\nu,\eta}^{-1}(\alpha_{i}),
    \label{Eq:VaR_estimation}
\end{equation}
with
\begin{equation}
    \mu_{i,t} = \varphi_{0} + \sum\limits_{j=1}^{m} \varphi_{j}r_{i,t-j} + \sum\limits_{j=1}^{n} \gamma_{j}\varepsilon_{i,t-j},
    \label{Eq:mu_estimation}
\end{equation}
\label{Eq:VaR_mu_estimation}%
\end{subequations}
where $t_{\nu,\eta}^{-1}(\alpha_{i})$ is the $\alpha_{i}$ quantile of the skewed Student-t distribution in Eq. (\ref{Eq:The_skewed_Student-t_density_distribution}). Specifically, the downside and upside $VaR$s for $r_{i}$ can be computed by Eq. (\ref{Eq:VaR_mu_estimation}) when $\alpha_{i}$ equals $\alpha_{i}^{d}$ and $\alpha_{i}^{u}$, respectively.

$CoVaR$ is adopted to quantify the extreme risk spillovers from the agricultural futures market to the corresponding spot market, the definition of which is the $\alpha_{2}$ quantile of the conditional distribution for spot returns conditional on the given $\alpha_{1}$ quantile of the conditional distribution for futures returns. Accordingly, the downside and upside $CoVaR$s for the spot return series $r_{2}$ are given by
\begin{subequations}
  \begin{equation}
    \text{Pr} \left( r_{2,t} \leq CoVaR_{\alpha_{2}^{d},\alpha_{1}^{d},t}^{r_{2,t} \mid r_{1,t}}  ~\Big|~ r_{1,t} \leq VaR_{\alpha_{1}^{d},t}^{r_{1,t}} \right) = \alpha_{2}^{d} 
    \label{Eq:Downside_CoVaR_estimation},
  \end{equation}
and
  \begin{equation}
    \text{Pr} \left( r_{2,t} \geq CoVaR_{\alpha_{2}^{u},\alpha_{1}^{u},t}^{r_{2,t} \mid r_{1,t}}  ~\Big|~ r_{1,t} \geq VaR_{\alpha_{1}^{u},t}^{r_{1,t}} \right) = \alpha_{2}^{u}, 
    \label{Eq:Upside_CoVaR_estimation}
  \end{equation}
\end{subequations}
where $\alpha_{1}^{d}$ and $\alpha_{1}^{u}$ are set as 0.05 and 0.95, respectively.

In combination with copula models, the $CoVaR$s for $r_{2}$ can be obtained by solving
\begin{equation}
    C \left( F_{2,t} \left( CoVaR_{\alpha_{2},\alpha_{1},t}^{r_{2,t} \mid r_{1,t}} \right), F_{1,t} \left( VaR_{\alpha_{1},t}^{r_{1,t}} \right) \right) - \alpha_{2}\alpha_{1}= 0, 
    \label{Eq:CoVaR_estimation_equation}
\end{equation}
where $F_{1,t}$ and $F_{2,t}$ refer to the respective marginal distribution functions of the futures and spot returns. Specifically, the downside $CoVaR$s for $r_{2}$ are calculated under conditions of $\alpha_{2}=\alpha_{2}^{d}$ and $\alpha_{1}=\alpha_{1}^{d}$, and the upside $CoVaR$s for $r_{2}$ are computed under conditions of $\alpha_{2}=\alpha_{2}^{u}$ and $\alpha_{1}=\alpha_{1}^{u}$.

Given the confidence levels $\alpha_{1}$ and $\alpha_{2}$, as well as the specification of the copula, we can invert the copula function in Eq. (\ref{Eq:CoVaR_estimation_equation}) to calculate the value of $F_{2,t} \left( CoVaR_{\alpha_{2},\alpha_{1},t}^{r_{2,t} \mid r_{1,t}} \right)$, and then invert the marginal distribution function of spot returns by $F_{2,t}^{-1} \left( F_{2,t} \left( CoVaR_{\alpha_{2},\alpha_{1},t}^{r_{2,t} \mid r_{1,t}} \right) \right)$ to obtain the value of $CoVaR_{\alpha_{2},\alpha_{1},t}^{r_{2,t} \mid r_{1,t}}$.

$\Delta CoVaR$ is also employed to quantify the extreme risk spillovers of the agricultural futures market to the corresponding spot market, which is defined as the change from the $VaR$ for spot returns conditional on an extreme movement of futures returns to the $VaR$ for spot returns conditional on a normal state of futures returns. Accordingly, the downside and upside $\Delta CoVaR$s for the spot return series $r_{2}$ are given by \begin{subequations}
  \begin{equation}
    \Delta CoVaR_{\alpha_{2}^{d},\alpha_{1},t}^{r_{2,t} \mid r_{1,t}} = CoVaR_{\alpha_{2}^{d},\alpha_{1} = 0.05,t}^{r_{2,t} \mid r_{1,t}} - CoVaR_{\alpha_{2}^{d},\alpha_{1} = 0.5,t}^{r_{2,t} \mid r_{1,t}}
    \label{Eq:Downside_delta_CoVaR_estimation}
  \end{equation}
and
  \begin{equation}
    \Delta CoVaR_{\alpha_{2}^{u},\alpha_{1},t}^{r_{2,t} \mid r_{1,t}} = CoVaR_{\alpha_{2}^{u},\alpha_{1} = 0.95,t}^{r_{2,t} \mid r_{1,t}} - CoVaR_{\alpha_{2}^{u},\alpha_{1} = 0.5,t}^{r_{2,t} \mid r_{1,t}},
    \label{Eq:Upside_delta_CoVaR_estimation}
  \end{equation}
\end{subequations}
where the value of the normal state is set as 0.5, and $CoVaR_{\alpha_{2}^{d},\alpha_{1} = 0.5,t}^{r_{2,t} \mid r_{1,t}}$ and $CoVaR_{\alpha_{2}^{u},\alpha_{1} = 0.5,t}^{r_{2,t} \mid r_{1,t}}$ satisfy
\begin{subequations}
  \begin{equation}
    \text{Pr} \left( r_{2,t} \leq CoVaR_{\alpha_{2}^{d},\alpha_{1} = 0.5,t}^{r_{2,t} \mid r_{1,t}}  ~\Big|~  F_{1,t}(r_{1,t}) = 0.5 \right) = \alpha_{2}^{d}
    \label{Eq:Downside_delta_CoVaR_estimation_requirement}
  \end{equation}
and
  \begin{equation}
    \text{Pr} \left( r_{2,t} \geq CoVaR_{\alpha_{2}^{u},\alpha_{1} = 0.5,t}^{r_{2,t} \mid r_{1,t}}  ~\Big|~  F_{1,t}(r_{1,t}) = 0.5 \right) = \alpha_{2}^{u}.
    \label{Eq:Upside_delta_CoVaR_estimation_requirement}
  \end{equation}
\end{subequations}

Furthermore, we refer to \cite{Reboredo-Ugolini-2015-JIntMoneyFinan, Reboredo-Ugolini-2016-EnergyEcon} and apply the K-S test proposed by \cite{Abadie-2002-JAmStatAssoc}, which can measure the discrepancy in the data, to assess the significance of extreme risk spillover effects, as well as the possible asymmetries of these extreme risk spillover effects at the directional and temporal levels. The K-S statistic can be defined as
\begin{equation}
    KS_{mn} = \left( \frac{mn}{m+n} \right)^{\frac{1}{2}} \mathop{\sup}_{x} \Big| G_{m}(x) - H_{n}(x) \Big|,
    \label{Eq:KS_test}
\end{equation}
where $G_{m}(x)$ denotes the empirical distribution
function of $CoVaR$ with $m$ samples, and $H_{n}(x)$ represents the empirical distribution
function of $VaR$ with $n$ samples.

\section{Data description}
\label{S1:Data}

\subsection{Data source}

We select the futures and spots of soybean, maize, wheat, and rice as our research objects, and utilize their daily price data from February 23, 2021, to February 24, 2023, to carry out our research. To obtain agricultural futures prices, we collect the daily closing prices of continuous contracts for soybean, corn, wheat, and rough rice futures listed on the Chicago Board of Trade (CBOT) from the Wind database. The CBOT futures prices serve as an important reference and benchmark for the international pricing of agricultural commodities. For agricultural spot prices, we gather daily data on price indices for soybean, maize, wheat, and rice from the official website of the International Grains Council (IGC). These price indices, derived from export quotations of multiple varieties, reliably reflect the dynamic trends in international agricultural spot prices. In addition, we determine the time node for the outbreak of the Russia-Ukraine conflict as February 24, 2022, and then divide the sample period into two sub-periods: February 23, 2021, to February 23, 2022 (pre-outbreak period), and February 24, 2022, to February 24, 2023 (post-outbreak period). This division enables us to investigate tail dependence structures and extreme risk spillovers between agricultural futures and spots before and after the outbreak of the Russia-Ukraine conflict, respectively.

\subsection{Statistical description}

\begin{figure}[!h]
  \centering
  \includegraphics[width=0.475\linewidth]{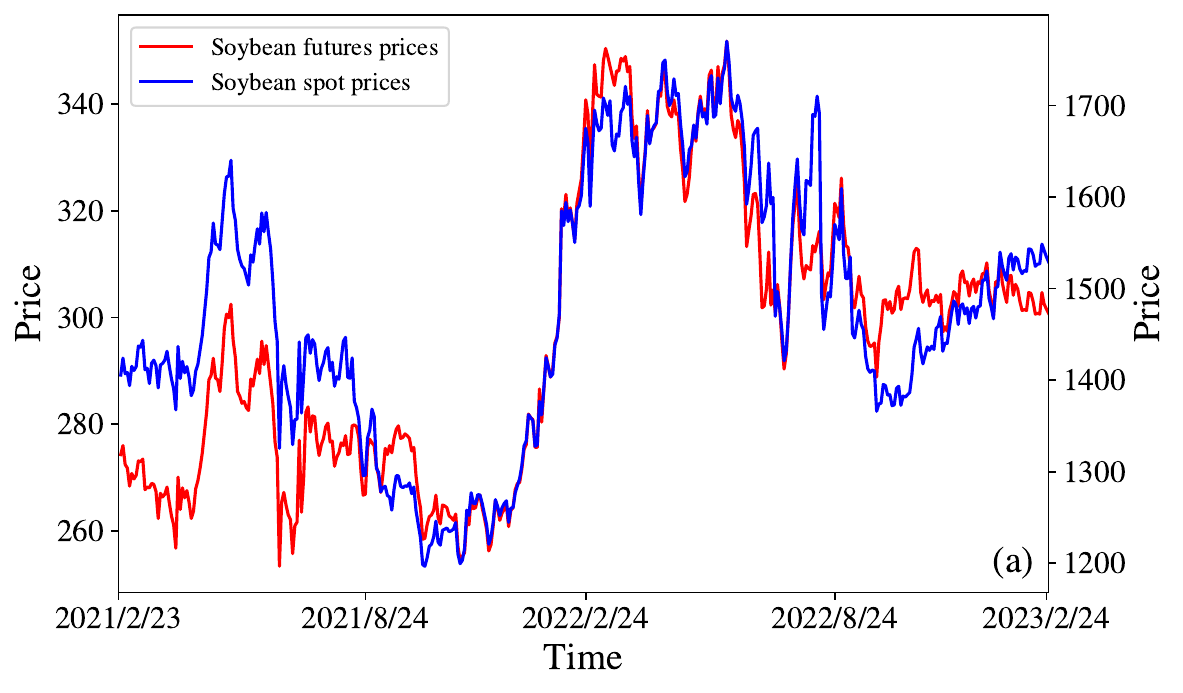}
  \includegraphics[width=0.475\linewidth]{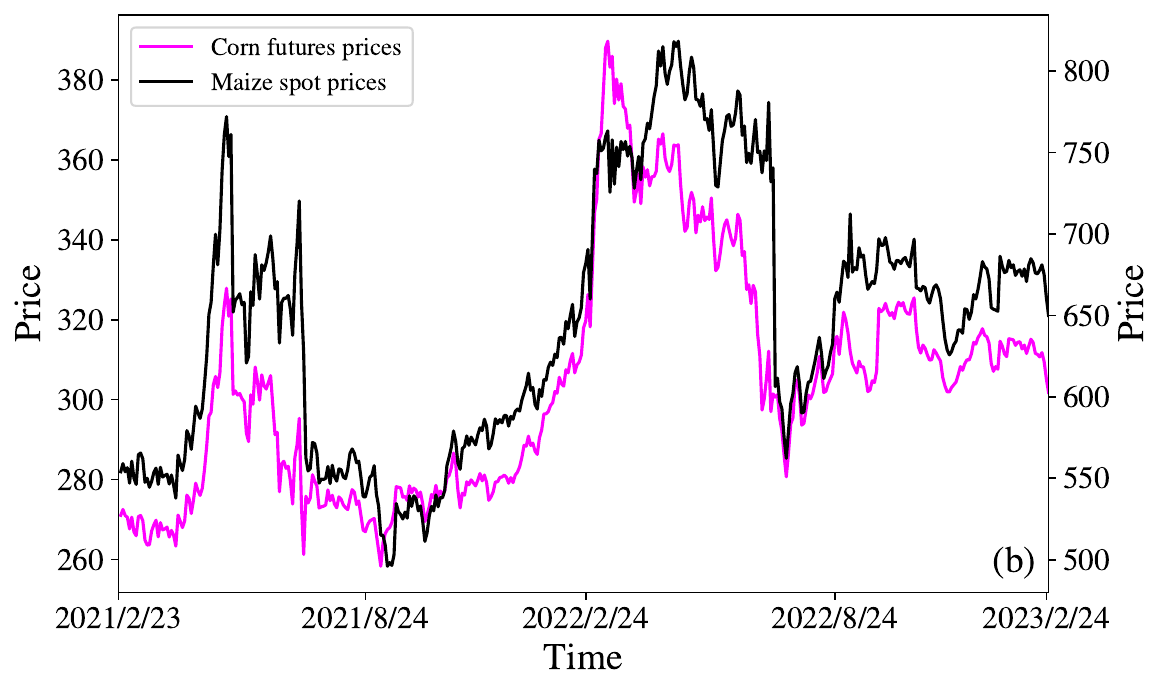}\\
  \includegraphics[width=0.475\linewidth]{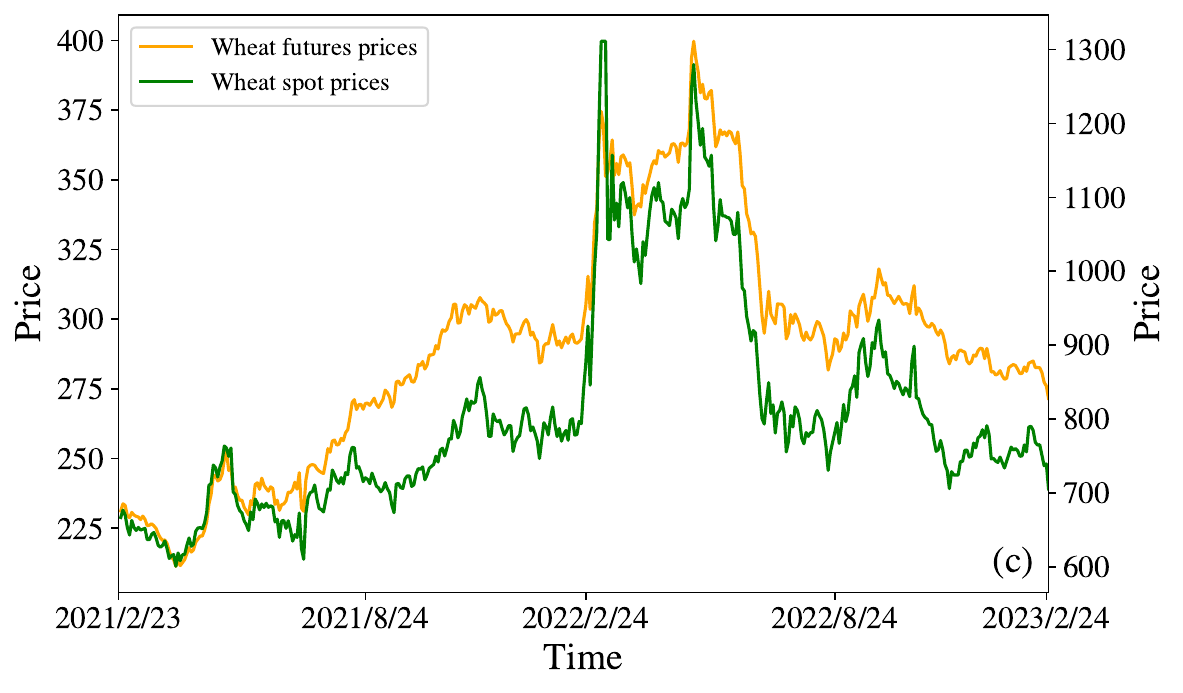}
  \includegraphics[width=0.475\linewidth]{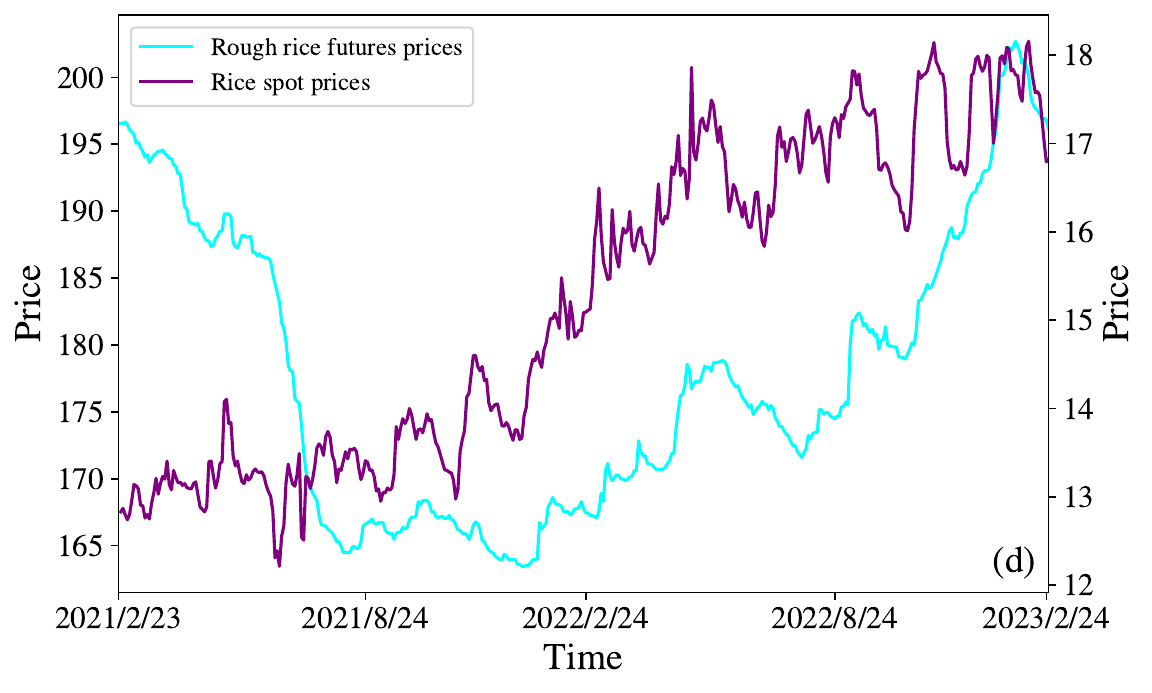}
  \caption{Price evolution of futures and spots of soybean (a), maize (b), wheat (c), and rice (d).}
\label{Fig:AgroPrice_evolution}
\end{figure}

Figure~\ref{Fig:AgroPrice_evolution} depicts the price movements of futures and spots for soybean, maize, wheat, and rice from February 23, 2021, to February 24, 2023. The prices of rough rice futures and rice spots exhibit an apparent discrepancy, especially during 2021, but their trends tend to be similar over time. By contrast, the overall trends between the futures and spot prices of soybean, maize, and wheat remain highly consistent, despite some short-term differentials. This finding suggests that the price series of the agricultural futures and spots selected in our paper can generally be regarded as a good match. Further analysis of Figure~\ref{Fig:AgroPrice_evolution} reveals that in April 2021, due to the anticipated decline in production and low inventories caused by the La Niña phenomenon, the futures and spot prices of soybean and maize experienced a sharp rise, along with an upward trend in the prices of wheat futures, wheat spots, and rice spots, and then swung downward in the following two months. Moreover, the futures and spot prices of the four grain crops have all reacted to the Russia-Ukraine conflict. Specifically, soybean and maize prices had shown sustained upward trends prior to the outbreak, peaking shortly after the conflict began. Wheat futures and spot prices surged after the outbreak, especially wheat spots, whose prices soared and peaked within a short period. Compared to the dramatic rises in soybean, maize, and wheat prices, rice prices demonstrated oscillating upward trends around the outbreak of the conflict, with a relatively limited magnitude. Additionally, the futures and spot prices of soybean, maize, and wheat have exhibited an overall downward trend since June 2022, whereas rice futures and spot prices have remained fluctuating upward.

Based on the alignment of daily data, we then calculate the logarithmic return of agricultural futures or spots at time $t$ and multiply it by 100, which is given by
\begin{equation}\label{Eq:Logarithmic_return}
   r_i(t) = \ln{\frac{P_i(t)}{P_i(t - \Delta t)}} \times 100
\end{equation}
where $\Delta t$ is set to 1 day, $i=1,2$, and $P_{1}(t)$, $P_{2}(t)$ refer to the daily prices of the agricultural futures and spots at time $t$, respectively.

Table~\ref{Tab:Correlation_Coefficient} provides the correlation coefficients between the futures and spot return series of soybean, maize, wheat, and rice for the whole sample period, as well as for the two sub-periods before and after the outbreak of the Russia-Ukraine conflict. The Pearson product-moment correlation coefficient, Kendall's tau coefficient, and Spearman's rank correlation coefficient are adopted as the three major statistical correlation coefficients to quantitatively describe the linear correlation between variables. As shown in Table~\ref{Tab:Correlation_Coefficient}, the correlation coefficients between the futures and spot return series of soybean, maize, and wheat are significantly positive across all periods, while the futures and spot return series of rice consistently exhibit a statistically insignificant positive correlation. Furthermore, except for the Kendall's tau and Spearman's rank correlation coefficients of wheat, the other correlation coefficients for the pre-outbreak period are all larger than their corresponding correlation coefficients for the post-outbreak period, which implies that the linear correlations between agricultural futures and spot returns have weakened to some extent after the outbreak of the Russia-Ukraine conflict.

\begin{table}[!ht]
  \centering
  \setlength{\abovecaptionskip}{0pt}
  \setlength{\belowcaptionskip}{10pt}
  \caption{Correlation coefficients between futures and spot return series of agricultural commodities} 
  \setlength\tabcolsep{11.5pt}
  \resizebox{\textwidth}{!}{ 
  \begin{tabular}{lcccc}
    \toprule
        & \multicolumn{1}{c}{Soybean} & \multicolumn{1}{c}{Maize} & \multicolumn{1}{c}{Wheat} & \multicolumn{1}{c}{Rice} \\
    \midrule
    \multicolumn{5}{l}{\textit{Panel A: In the whole sample period}} \\
    Pearson product-moment correlation coefficient & 0.8851$^{***}$ & 0.7222$^{***}$ & 0.8128$^{***}$ & 0.0487 \\
    Kendall's tau coefficient & 0.8196$^{***}$ & 0.7265$^{***}$ & 0.6959$^{***}$ & 0.0491 \\
    Spearman's rank correlation coefficient & 0.9411$^{***}$ & 0.8722$^{***}$ & 0.8668$^{***}$ & 0.0698  \vspace{2mm}\\
    \multicolumn{5}{l}{\textit{Panel B: In the pre-outbreak period}} \\
    Pearson product-moment correlation coefficient & 0.9535$^{***}$ & 0.7818$^{***}$ & 0.8725$^{***}$ & 0.1111 \\
    Kendall's tau coefficient & 0.8339$^{***}$ & 0.7623$^{***}$ & 0.6642$^{***}$ & 0.0526 \\
    Spearman's rank correlation coefficient & 0.9498$^{***}$ & 0.8889$^{***}$ & 0.8425$^{***}$ & 0.0783  \vspace{2mm}\\
    \multicolumn{5}{l}{\textit{Panel C: In the post-outbreak period}} \\
    Pearson product-moment correlation coefficient & 0.8246$^{***}$ & 0.6560$^{***}$ & 0.7898$^{***}$ & 0.0098 \\
    Kendall's tau coefficient & 0.8053$^{***}$ & 0.7004$^{***}$ & 0.7262$^{***}$ & 0.0477 \\
    Spearman's rank correlation coefficient & 0.9321$^{***}$ & 0.8595$^{***}$ & 0.8843$^{***}$ & 0.0660 \\
    \bottomrule
  \end{tabular}} 
  \begin{flushleft}
    \footnotesize
    \justifying Note: This table reports the results of the three major statistical correlation coefficients, which can be adopted to quantitatively describe the linear correlation between the futures and spot return series of soybean, maize, wheat, and rice over the whole sample period and over the two sub-periods before and after the outbreak of the Russia-Ukraine conflict. These correlation coefficients take values in the interval $[-1,1]$, where $-$1 and 1 denote perfectly negative and perfectly positive correlations, respectively, and 0 corresponds to no correlation. The superscript $^{***}$ indicates significance at the 1\% level, and nonsubscripted coefficients represent insignificance at the 10\% level.
 \end{flushleft} 
  \label{Tab:Correlation_Coefficient}%
\end{table}%



Table~\ref{Tab:Agro_Stat_Test} reports the descriptive statistics of the futures and spot return series of soybean, maize, wheat, and rice for the whole sample period, as well as for the two sub-periods before and after the outbreak of the Russia-Ukraine conflict. Comparing the results, we observe that each spot return series has a larger maximum value, standard deviation, and absolute minimum value than its corresponding futures return series over various periods, indicating a wider fluctuation range of agricultural spot returns. Additionally, the skewness of each return series deviates from zero, which means that the distribution of each agricultural return series is left- or right-skewed. In Panel A of Table~\ref{Tab:Agro_Stat_Test}, the kurtosis of each return series for the whole sample period exceeds that of a normal distribution, confirming the presence of leptokurtosis and fat tails in the agricultural return series. As evidenced by Panels B and C, except for the wheat spots for the pre-outbreak period and the soybean and corn futures for the post-outbreak period, the other return series over the two sub-periods also exhibit the characteristics of leptokurtosis and fat tails.

Panel A of Table~\ref{Tab:Agro_Stat_Test} further provides the diagnostic test results for the futures and spot return series of soybean, maize, wheat, and rice for the whole sample period, including the normality, unit root, white noise, and ARCH effect tests. As can be seen, all the Jarque-Bera statistics for the return series are significantly larger than 0, implying that the futures-spot return series of each agricultural commodity do not obey the normal distribution. The significant statistics of the ADF and PP tests, along with the insignificant statistic of the KPSS test, consistently suggest that each agricultural return series is stationary. Moreover, the Q and Q$^{2}$ statistics obtained from the Ljung-Box test refer to the results of the white noise test on the return and squared return series, respectively, indicating the presence of serial autocorrelation and conditional heteroskedasticity in each agricultural return series. In addition, the significant statistics of the ARCH-LM test further confirm the existence of ARCH effects in all the return series. Based on the above analysis, we therefore construct the ARMA-GARCH-skewed Student-t model for the futures-spot return series of soybean, maize, wheat, and rice to conduct further research.

\begin{table}[!ht]
  \centering
  \setlength{\abovecaptionskip}{0pt}
  \setlength{\belowcaptionskip}{10pt}
  \caption{Descriptive statistics and diagnostic tests for futures and spot return series of agricultural commodities}
  \setlength\tabcolsep{3pt}   \resizebox{\textwidth}{!}{ 
    \begin{tabular}{l r@{.}l r@{.}l r@{.}l r@{.}l c r@{.}l r@{.}l r@{.}l r@{.}l}
    \toprule
         & \multicolumn{8}{c}{Futures} && \multicolumn{8}{c}{Spots}  \\
    \cline{2-9} \cline{11-18}
         & \multicolumn{2}{c}{Soybean} & \multicolumn{2}{c}{Corn} & \multicolumn{2}{c}{Wheat} & \multicolumn{2}{c}{Rough rice} && \multicolumn{2}{c}{Soybean} & \multicolumn{2}{c}{Maize} & \multicolumn{2}{c}{Wheat} & \multicolumn{2}{c}{Rice}  \\
    \midrule
    \multicolumn{18}{l}{\textit{Panel A: In the whole sample period}} \\
    Max & 6&5886 & 5&4336 & 6&7993 & 2&5371 && 7&3975 & 6&1683 & 13&1690 & 7&3161 \\
    Min & $-$7&7505 & $-$7&7492 & $-$5&3321 & $-$1&1580 && $-$12&4326 & $-$20&0033 & $-$22&8880 & $-$7&2626 \\
    Mean & 0&0219 & 0&0256 & 0&0382 & $-$0&0004 && 0&0200 & 0&0380 & 0&0134 & 0&0643 \\
    Std. Dev. & 1&4562 & 1&4684 & 1&3732 & 0&3481 && 1&8343 & 2&3221 & 2&9007 & 1&3860 \\
    Skewness & 0&0307 & $-$0&5678 & 0&3941 & 1&5390 && $-$1&0278 & $-$2&4763 & $-$0&6418 & 0&3993 \\
    Kurtosis & 3&2865 & 4&5700 & 4&1686 & 9&7718 && 7&1457 & 18&3966 & 10&8636 & 4&9608 \vspace{1mm}\\
    Jarque-Bera & 193&0100$^{***}$ & 394&9500$^{***}$ & 320&8300$^{***}$ & 1863&3000$^{***}$ && 982&5100$^{***}$ & 6436&1000$^{***}$ & 2124&6000$^{***}$ & 449&5900$^{***}$  \\
    ADF         & $-$6&7545$^{***}$ & $-$6&0670$^{***}$ & $-$6&2671$^{***}$ & $-$5&7347$^{***}$ && $-$7&5514$^{***}$ & $-$6&9566$^{***}$ & $-$6&4515$^{***}$ & $-$7&8556$^{***}$  \\
    PP          & $-$408&5400$^{***}$ & $-$401&2800$^{***}$ & $-$358&0800$^{***}$ & $-$384&3300$^{***}$ && $-$398&3800$^{***}$ & $-$454&7700$^{***}$ & $-$393&9500$^{***}$ & $-$336&2600$^{***}$  \\
    KPSS        & 0&0829 & 0&1096 & 0&1545 & 0&1662 & & 0&0695 & 0&0748 & 0&1475 & 0&0455  \\
    Q(15)   & 27&4530$^{**}$ & 20&6960$^{*}$ & 24&0700$^{*}$ & 165&3700$^{***}$ && 26&6910$^{**}$ & 33&8080$^{***}$ & 25&3580$^{**}$ & 27&0430$^{**}$  \\
    Q(20)       & 41&7130$^{***}$ & 26&5930$^{*}$ & 29&8050$^{*}$ & 168&8800$^{***}$ && 36&8550$^{**}$ & 37&9840$^{***}$ & 31&4510$^{**}$ & 32&8160$^{**}$  \\
    Q$^{2}$(15) & 69&1960$^{***}$ & 71&2140$^{***}$ & 118&6200$^{***}$ & 23&7120$^{*}$ && 35&6120$^{***}$ & 22&6760$^{*}$ & 100&1500$^{***}$ & 25&2570$^{**}$  \\
    Q$^{2}$(20) & 71&2360$^{***}$ & 73&2450$^{***}$ & 120&4400$^{***}$ & 26&8520$^{*}$ && 37&1340$^{**}$ & 24&4640$^{*}$ & 100&4300$^{***}$ & 28&7060$^{*}$  \\
    ARCH-LM(15) & 43&6350$^{***}$ & 40&3960$^{***}$ & 74&2130$^{***}$ & 21&9160$^{*}$ && 27&4990$^{**}$ & 20&6690$^{*}$ & 57&5880$^{***}$ & 24&0210$^{*}$  \vspace{2mm}\\
    \multicolumn{18}{l}{\textit{Panel B: In the pre-outbreak period}} \\
    Max & 6&5886 & 5&4336 & 4&7365 & 1&6719 && 7&3975 & 6&1683 & 9&6390 & 5&5959 \\
    Min & $-$7&7505 & $-$7&7492 & $-$5&3321 & $-$1&1580 &&  $-$8&4588 & $-$15&4589 & $-$8&1084 & $-$7&2626 \\
    Mean & 0&1029 & 0&0782 & 0&1307 & $-$0&0759 && 0&0831 & 0&0983 & 0&1285 & 0&0769 \\
    Std. Dev. & 1&5680 & 1&5064 & 1&1253 & 0&2949 && 1&7580 & 2&3739 & 2&1300 & 1&3291 \\
    Skewness & 0&1611 & $-$1&1302 & $-$0&1717 & 0&1654 && $-$0&0755 & $-$1&8662 & 0&2322 & 0&1142 \\
    Kurtosis & 4&7014 & 7&0045 & 3&3940 & 7&7470 && 3&9694 & 10&5910 & 2&5621 & 6&3902  \vspace{2mm}\\ 
    \multicolumn{18}{l}{\textit{Panel C: In the post-outbreak period}} \\
    Max & 3&0415 & 4&9793 & 6&7993 & 2&5371 && 4&6943 & 5&9527 & 13&1690 & 7&3161 \\
    Min & $-$3&7523 & $-$4&9789 & $-$5&1336 & $-$0&8059 && $-$12&4326 & $-$20&0033 & $-$22&8880 & $-$5&5250 \\
    Mean & $-$0&0595 & $-$0&0274 & $-$0&0548 & 0&0754 && $-$0&0434 & $-$0&0227 & $-$0&1023 & 0&0517 \\
    Std. Dev. & 1&3334 & 1&4308 & 1&5811 & 0&3803 && 1&9101 & 2&2730 & 3&5105 & 1&4441 \\
    Skewness & $-$0&2528 & 0&0830 & 0&6665 & 2&0517 && $-$1&7507 & $-$3&1645 & $-$0&7159 & 0&6223 \\
    Kurtosis & $-$1&7463 & 1&6732 & 3&6696 & 9&1165 && 9&0387 & 27&3596 & 9&0891 & 3&7924  \\
  \bottomrule
    \end{tabular}
    }%
  \begin{flushleft}
    \footnotesize
    \justifying Note: This table reports the descriptive statistics and diagnostic test results for each agricultural return series throughout the sample period, as well as the descriptive statistics for return series before and after the outbreak of the Russia-Ukraine conflict, where the returns are calculated by multiplying the logarithmic returns by 100. The Jarque-Bera test is a normality test, where a statistic much larger than 0 indicates a significant departure from normality. The ADF, PP and KPSS tests are unit root tests, and the null hypothesis in both ADF test and PP test is non-stationarity, while the null hypothesis in KPSS test is stationarity. The Q and Q$^{2}$ statistics for the Ljung-Box test appertain to the results of the white noise test on the return and squared return series, respectively. The ARCH-LM test examines the presence of ARCH effect. $^{***}$, $^{**}$ and $^{*}$ denote significance at the 1\%, 5\% and 10\% level, respectively.
\end{flushleft} 
  \label{Tab:Agro_Stat_Test}%
\end{table}%

\section{Empirical analysis}
\label{S1:EmpAnal}

\subsection{Estimation of marginal models}

Table~\ref{Tab:Agro_Marginal_estimation} reports the marginal parameter estimates based on the ARMA-GARCH-skewed Student-t models for the futures and spot return series of soybean, maize, wheat, and rice, respectively, with the standard errors of the estimated coefficients listed in parentheses. According to the AIC, we determine the optimal combination of lag parameters $m$, $n$, $p$, and $q$ for each marginal model in the range of 0 to 3. From Panel A of Table~\ref{Tab:Agro_Marginal_estimation}, it can be found that the mean equations for different return series follow various ARMA$(m, n)$ models, with the statistically significant coefficients at the 1\% level except for some constant terms. In Panel B, the variance equations for the agricultural return series involve diverse ARCH and GARCH terms, most of which are statistically significant. Additionally, the estimates of the asymmetric and degrees-of-freedom parameters for each series are both significant at the 1\% level, indicating that the error terms of the marginal model for each agricultural return series conform to a right-skewed distribution with fat tails. Hence, the skewed Student-t model is capable of accurately describing the distribution characteristics of the error terms.

\begin{table}[!ht]
  \centering
  \setlength{\abovecaptionskip}{0pt}
  \setlength{\belowcaptionskip}{10pt}
  \caption{Marginal estimation using the ARMA-GARCH-skewed Student-t model presented in Eq.~(\ref{Eq:Marginal_distribution})}
    \setlength\tabcolsep{2.9pt}
    \resizebox{\textwidth}{!}{
    \begin{tabular}{l r@{.}l r@{.}l r@{.}l r@{.}l c r@{.}l r@{.}l r@{.}l r@{.}l}
    \toprule
    & \multicolumn{8}{c}{Futures} && \multicolumn{8}{c}{Spots} \\
    \cline{2-9} \cline{11-18}
    & \multicolumn{2}{c}{Soybean} & \multicolumn{2}{c}{Corn} & \multicolumn{2}{c}{Wheat} & \multicolumn{2}{c}{Rough rice} && \multicolumn{2}{c}{Soybean} & \multicolumn{2}{c}{Maize} & \multicolumn{2}{c}{Wheat} & \multicolumn{2}{c}{Rice} \\
    \midrule
    \multicolumn{18}{l}{\textit{Panel A: Mean equation in Eq.~(\ref{Eq:Marginal_distribution_return})}} \\
    Constant, $\varphi_0$ & 0&1048$^{***}$ & 0&0318 & 0&0123$^{***}$ & $-$0&0206$^{***}$ && 0&0185 & 0&0994$^{***}$ & $-$0&0041$^{***}$ & 0&0609$^{***}$ \\
         & (0&0001) & (0&0588) & (0&0001) & (0&0026) && (0&0824) & (0&0135) & (0&0000) & (0&0000) \\
    AR(1), $\varphi_1$ & 1&7832$^{***}$ & $-$0&0443$^{***}$ & 1&9801$^{***}$ & 2&6590$^{***}$ && $-$0&8909$^{***}$ & 2&7096$^{***}$ & $-$0&7817$^{***}$ & 0&9405$^{***}$ \\
         & (0&0001) & (0&0164) & (0&0004) & (0&0008) && (0&0515) & (0&0002) & (0&0132) & (0&0205) \\
    AR(2), $\varphi_2$ & $-$1&1811$^{***}$ & $-$0&0664$^{***}$ & $-$0&9944$^{***}$ & $-$2&5038$^{***}$ && $-$0&7809$^{***}$ & $-$2&6222$^{***}$ & 0&8824$^{***}$ & \multicolumn{2}{r}{} \\
         & (0&0001) & (0&0146) & (0&0002) & (0&0008) && (0&0481) & (0&0002) & (0&0001) & \multicolumn{2}{r}{} \\
    AR(3), $\varphi_3$ & 0&1216$^{***}$ & $-$0&9385$^{***}$ & \multicolumn{2}{r}{} & 0&8168$^{***}$ && \multicolumn{2}{r}{} & 0&8918$^{***}$ & 0&8226$^{***}$ & \multicolumn{2}{r}{} \\
         & (0&0000) & (0&0142) & \multicolumn{2}{r}{} & (0&0004) && \multicolumn{2}{r}{} & (0&0001) & (0&0002) & \multicolumn{2}{r}{} \\
    MA(1), $\gamma_1$ & $-$1&7538$^{***}$ & 0&0467$^{***}$ & $-$1&9244$^{***}$ & $-$2&5235$^{***}$ && 0&9800$^{***}$ & $-$2&7370$^{***}$ & 0&8273$^{***}$ & $-$0&7961$^{***}$ \\
         & (0&0001) & (0&0009) & (0&0002) & (0&0010) && (0&0291) & (0&0003) & (0&0001) & (0&0003) \\
    MA(2), $\gamma_2$ & 1&0527$^{***}$ & 0&1210$^{***}$ & 0&8391$^{***}$ & 2&2519$^{***}$ && 0&9238$^{***}$ & 2&6784$^{***}$ & $-$0&9571$^{***}$ & $-$0&2351$^{***}$ \\
         & (0&0001) & (0&0014) & (0&0003) & (0&0010) && (0&0271) & (0&0003) & (0&0001) & (0&0001) \\
    MA(3), $\gamma_3$ & \multicolumn{2}{r}{} & 0&9773$^{***}$ & 0&1020$^{***}$ & $-$0&6782$^{***}$ && \multicolumn{2}{r}{} & $-$0&9199$^{***}$ & $-$0&9552$^{***}$ & \multicolumn{2}{r}{} \\
         & \multicolumn{2}{r}{} & (0&0005) & (0&0000) & (0&0002) && \multicolumn{2}{r}{} & (0&0001) & (0&0001) & \multicolumn{2}{r}{} \vspace{2mm} \\
    \multicolumn{18}{l}{\textit{Panel B: Variance equation in Eq.~(\ref{Eq:Marginal_distribution_conditional_variance})}} \\
    Constant, $\alpha_0$ & 1&2874$^{***}$ & 0&0809 & 0&8635$^{***}$ & 0&1850 && 0&2735$^{*}$ & 1&9219$^{***}$ & 1&2649$^{*}$ & 1&2868$^{***}$ \\
           & (0&1170) & (0&0610) & (0&1444) & (0&1280) && (0&1469) & (0&4142) & (0&7442) & (0&3174) \\ 
    ARCH(1), $\alpha_1$ & 0&2352$^{***}$ & 0&1614 & 0&2130$^{**}$ & 0&9990$^{**}$ && 0&1051$^{**}$ & 0&3557$^{*}$ & 0&2684$^{**}$ & 0&2691$^{*}$ \\
         & (0&0776) & (0&1056) & (0&0878) & (0&4585) && (0&0444) & (0&2008) & (0&1330) & (0&1404) \\
    ARCH(2), $\alpha_2$ & \multicolumn{2}{r}{} & 0&0020 & 0&3389$^{***}$ & \multicolumn{2}{r}{} && \multicolumn{2}{r}{} & 0&4631$^{**}$ & \multicolumn{2}{r}{} & 0&1014 \\
         & \multicolumn{2}{r}{} & (0&1267) & (0&1110) & \multicolumn{2}{r}{} && \multicolumn{2}{r}{} &	(0&2099) & \multicolumn{2}{r}{} & (0&0740) \\
    GARCH(1), $\beta_1$ & \multicolumn{2}{r}{} & 0&8119$^{***}$ & \multicolumn{2}{r}{} & 	\multicolumn{2}{r}{} &&	0&8125$^{***}$ & \multicolumn{2}{r}{} & 0&5496$^{*}$ & \multicolumn{2}{r}{} \\
         & \multicolumn{2}{r}{} & (0&0868) & \multicolumn{2}{r}{} & \multicolumn{2}{r}{} && (0&0679) & \multicolumn{2}{r}{} & (0&3144) & \multicolumn{2}{r}{} \\
    GARCH(2), $\beta_2$ & \multicolumn{2}{r}{} & \multicolumn{2}{r}{} & \multicolumn{2}{r}{} & \multicolumn{2}{r}{} && \multicolumn{2}{r}{} & \multicolumn{2}{r}{} & 0&0425 & \multicolumn{2}{r}{} \\ 
          & \multicolumn{2}{r}{} & \multicolumn{2}{r}{} & \multicolumn{2}{r}{} & \multicolumn{2}{r}{} && \multicolumn{2}{r}{} & \multicolumn{2}{r}{} & (0&2163) & \multicolumn{2}{r}{} \\
    Asymmetry, $\eta$ & 0&9844$^{***}$ & 0&9537$^{***}$ & 0&9769$^{***}$ & 1&0003$^{***}$ && 0&8547$^{***}$ & 0&8842$^{***}$ & 1&0134$^{***}$ & 1&1836$^{***}$ \\
          & (0&0603) & (0&0659) & (0&0583) & (0&0399) && (0&0611) & (0&0592) & (0&0680) & (0&0767) \\
    Tail, $\nu$ & 33&1938$^{***}$ & 4&6875$^{***}$ & 4&8303$^{***}$ & 2&2098$^{***}$ && 4&3771$^{***}$ & 4&4198$^{***}$ & 4&9146$^{***}$ & 3&6836$^{***}$ \\
        & (11&1204) & (1&1306) & (1&1886) & (0&1664) && (0&9544) & (1&5992) & (1&5715) & (0&8383)  \vspace{2mm}\\
    \multicolumn{18}{l}{\textit{Panel C: Diagnostic tests}} \\
    LL & \multicolumn{2}{c}{$-$712.00} & \multicolumn{2}{c}{$-$684.11} & \multicolumn{2}{c}{$-$655.74} & \multicolumn{2}{c}{$-$9.54} && \multicolumn{2}{c}{$-$789.08} & \multicolumn{2}{c}{$-$847.24} & \multicolumn{2}{c}{$-$955.38} & \multicolumn{2}{c}{$-$661.16} \\	 	
    AIC & 3&4299 & 3&3117 & 3&1674 & 0&0976 && 3&7961 & 4&0819 & 4&6004 & 3&1837 \\
    Q(15) & [0&1207] & [0&9213] & [0&4959] & [0&2680] && [0&4567] & [0&1651] & [0&4696] & [0&4055] \\
    Q(20) & [0&1122] & [0&5684] & [0&2865] & [0&1573] && [0&4201] & [0&1057] & [0&3334] & [0&4648] \\
    Q$^{2}$(15) & [0&2931] & [0&9987] & [0&9398] & [0&9688] && [0&7676] & [0&6493] & [0&9909] & [0&8681] \\	
    Q$^{2}$(20) & [0&3881] & [0&9946] & [0&9782] & [0&9911] && [0&8730] & [0&8010] & [0&9983] & [0&9467] \\
    ARCH-LM(15) & [0&1065] & [0&9990] & [0&9570] & [0&9731] && [0&8803] & [0&7174] & [0&9915] & [0&9367] \\
    ARCH-LM(20) & [0&1076] & [0&9973] & [0&9883] & [0&9931] && [0&9501] & [0&8421] & [0&9984] & [0&9825] \\
   \bottomrule
    \end{tabular}
    }%
  \begin{flushleft}
    \footnotesize
    \justifying Note: This table presents the estimated results and diagnostic tests for each marginal ARMA-GARCH-skewed Student-t model, in which the optimal lag parameters are determined in the range of 0 to 3 according to the AIC. The Jarque-Bera test, Ljung-Box test and ARCH-LM test are utilized to examine the existence of normality, serial correlation and ARCH effect in the standardized residual sequence for each marginal model, respectively. The standard errors of parameter estimates are listed in parentheses, and the $p$-values of test statistics are reported in square brackets. $^{***}$, $^{**}$ and $^{*}$ denote significance at the 1\%, 5\% and 10\% level, respectively.
  \end{flushleft}
  \label{Tab:Agro_Marginal_estimation}%
\end{table}%

Panel C provides the values of the logarithmic likelihood and Akaike information criterion for each marginal model, and the p-values corresponding to the diagnostic test statistics for each standardized residual sequence, which are listed in square brackets. As can be seen, the Q and Q$^{2}$ statistics of the Ljung-Box test are insignificant even at the 10\% level, suggesting the absence of serial autocorrelation and conditional heteroscedasticity in each standardized residual sequence. The results of the ARCH-LM test also verify the absence of ARCH effects in all standardized residual sequences, which aligns with our previous findings. By comparing the test results shown in Tables~\ref{Tab:Agro_Marginal_estimation} and \ref{Tab:Agro_Stat_Test}, we conclude that the marginal distribution of each agricultural return series is specified well by the ARMA-GARCH-skewed Student-t model constructed in our paper, providing an essential prerequisite for the parameter estimation of the copula models in the following empirical analysis.


\subsection{Estimation of single copula models}

With the marginal parameter estimates, we next estimate the parameters of copula models for each agricultural futures-spot pair. Considering that agricultural futures and spot returns are positively correlated over the whole sample period, as well as over the two sub-periods before and after the outbreak of the Russia-Ukraine conflict, we select six different types of single copula models, namely the Normal, Student-t, Clayton, survival Clayton, Gumbel, and survival Gumbel copulas, to evaluate the tail dependence between the agricultural futures and spot markets.

Table~\ref{Tab:Agro_Single_Copula_Estimation} provides the estimated results of the single copula models for the futures-spot pairs of soybean, maize, wheat, and rice before and after the outbreak of the Russia-Ukraine conflict, respectively, and the standard errors of the estimated coefficients are provided within parentheses. As can be seen, except for the rice futures-spot pair, the estimates of various single copula models for the other agricultural pairs are all significant at the 5\% level. Moreover, based on a comparison of the results, we notice that, for each single copula model regarding each specific agricultural futures-spot pair, the copula parameter estimate for the pre-outbreak period is larger than that for the post-outbreak period. For instance, as shown in Panel B, the parameter estimate of the Student-t copula for maize is 0.8768 before the outbreak of the conflict, while the corresponding estimate is 0.8371 afterward. This finding seems to indicate that the positive dependence between the futures and spot returns of each agricultural commodity has weakened to some extent after the outbreak of the Russia-Ukraine conflict, which is consistent with the changes in the correlation coefficients shown in Table~\ref{Tab:Correlation_Coefficient}. In other words, there appears to be less co-movement between the international agricultural futures and spot markets during the post-outbreak period than during the pre-outbreak period.

\begin{table}[!ht]
  \centering
  \setlength{\abovecaptionskip}{0pt}
  \setlength{\belowcaptionskip}{10pt}
  \caption{Estimation of single copula models between futures and spot return series of agricultural commodities before and after the outbreak of the Russia-Ukraine conflict}
  \setlength\tabcolsep{3.3pt}
    \begin{tabular}{l r@{.}l r@{.}l c r@{.}l r@{.}l c r@{.}l  r@{.}l c r@{.}l r@{.}l}
    \toprule
        & \multicolumn{4}{c}{Soybean} && \multicolumn{4}{c}{Maize} && \multicolumn{4}{c}{Wheat} && \multicolumn{4}{c}{Rice} \\
        \cline{2-5}\cline{7-10}\cline{12-15}\cline{17-20}
        & \multicolumn{2}{c}{Before} & \multicolumn{2}{c}{After} && \multicolumn{2}{c}{Before} & \multicolumn{2}{c}{After} && \multicolumn{2}{c}{Before} & \multicolumn{2}{c}{After} && \multicolumn{2}{c}{Before} & \multicolumn{2}{c}{After}  \\
    \midrule
    \multicolumn{20}{l}{\textit{Panel A: Normal copula}} \\
    $\rho$ & \multicolumn{2}{c}{\textbf{0.8886}} & \multicolumn{2}{c}{\textbf{0.8478}} && \multicolumn{2}{c}{\textbf{0.8567}} & \multicolumn{2}{c}{\textbf{0.8109}} && \multicolumn{2}{c}{\textbf{0.8161}} & \multicolumn{2}{c}{\textbf{0.7944}} && \multicolumn{2}{c}{0.1438} & \multicolumn{2}{c}{0.0385} \\
        & \multicolumn{2}{c}{(0.0108)} & \multicolumn{2}{c}{(0.0149)} && \multicolumn{2}{c}{(0.0139)} & \multicolumn{2}{c}{(0.0185)} && \multicolumn{2}{c}{(0.0181)} & \multicolumn{2}{c}{(0.0197)} && \multicolumn{2}{c}{(0.0748)} & \multicolumn{2}{c}{(0.0651)} \\
    AIC & $-$326&9779 & $-$255&7947 && $-$278&5285 & $-$213&7208 && $-$206&6291  & $-$221&3889 && $-$1&4828 & 1&6507 \vspace{2mm} \\
    \multicolumn{20}{l}{\textit{Panel B: Student-t copula}} \\
    $\rho$ & \multicolumn{2}{c}{\textbf{0.8951}} & \multicolumn{2}{c}{\textbf{0.8587}} && \multicolumn{2}{c}{\textbf{0.8768}} & \multicolumn{2}{c}{\textbf{0.8371}} && \multicolumn{2}{c}{\textbf{0.8162}} & \multicolumn{2}{c}{\textbf{0.8043}} && \multicolumn{2}{c}{0.1198} & \multicolumn{2}{c}{0.0600} \\
        & \multicolumn{2}{c}{(0.0127)} & \multicolumn{2}{c}{(0.0173)} && \multicolumn{2}{c}{(0.0166)} & \multicolumn{2}{c}{(0.0204)} && \multicolumn{2}{c}{(0.0206)} & \multicolumn{2}{c}{(0.0225)} && \multicolumn{2}{c}{(0.0828)} & \multicolumn{2}{c}{(0.0711)} \\
    Dof $\nu$ & \multicolumn{2}{c}{\textbf{5.9800}} & \multicolumn{2}{c}{\textbf{5.1266}} && \multicolumn{2}{c}{\textbf{3.6773}} & \multicolumn{2}{c}{\textbf{4.5669}} && \multicolumn{2}{c}{\textbf{10.5215}} & \multicolumn{2}{c}{\textbf{6.9412}} && \multicolumn{2}{c}{5.7495} & \multicolumn{2}{c}{7.7833} \\
        & \multicolumn{2}{c}{(2.4121)} & \multicolumn{2}{c}{(2.0803)} && \multicolumn{2}{c}{(1.1404)} & \multicolumn{2}{c}{(1.5734)} && \multicolumn{2}{c}{(8.9893)} & \multicolumn{2}{c}{(3.0010)} && \multicolumn{2}{c}{(2.8575)} & \multicolumn{2}{c}{(5.1750)} \\
    AIC & $-$339&1537$^{\#}$ & $-$264&1735$^{\#}$ && $-$314&1136$^{\#}$ & $-$240&1268$^{\#}$ && $-$206&7270$^{\#}$ & $-$231&5107$^{\#}$ && $-$4&1397 & 1&0290$^{\#}$ \vspace{2mm}\\
    \multicolumn{20}{l}{\textit{Panel C: Clayton copula}} \\
    $\alpha$ & \multicolumn{2}{c}{\textbf{3.0208}} & \multicolumn{2}{c}{\textbf{2.5770}} && \multicolumn{2}{c}{\textbf{2.8880}} & \multicolumn{2}{c}{\textbf{2.2229}} && \multicolumn{2}{c}{\textbf{2.0289}} & \multicolumn{2}{c}{\textbf{1.7387}} && \multicolumn{2}{c}{0.1567} & \multicolumn{2}{c}{0.0170} \\
             & \multicolumn{2}{c}{(0.2503)} & \multicolumn{2}{c}{(0.2276)} && \multicolumn{2}{c}{(0.2491)} & \multicolumn{2}{c}{(0.2055)} && \multicolumn{2}{c}{(0.1957)} & \multicolumn{2}{c}{(0.1727)} && \multicolumn{2}{c}{(0.0886)} & \multicolumn{2}{c}{(0.0822)} \\
    AIC & $-$258&8400 & $-$225&0162 && $-$245&6114 & $-$191&7595 && $-$162&3030 & $-$165&6194 && $-$2&2948 & 1&9558 \vspace{2mm}\\
    \multicolumn{20}{l}{\textit{Panel D: Survival Clayton copula}} \\
    $\alpha$ & \multicolumn{2}{c}{\textbf{2.8877}} & \multicolumn{2}{c}{\textbf{2.2107}} && \multicolumn{2}{c}{\textbf{2.4945}} & \multicolumn{2}{c}{\textbf{2.1018}} && \multicolumn{2}{c}{\textbf{2.0131}} & \multicolumn{2}{c}{\textbf{1.9438}} && \multicolumn{2}{c}{0.1814} & \multicolumn{2}{c}{0.0650} \\
        & \multicolumn{2}{c}{(0.2437)} & \multicolumn{2}{c}{(0.2102)} && \multicolumn{2}{c}{(0.2201)} & \multicolumn{2}{c}{(0.2072)} && \multicolumn{2}{c}{(0.1930)} & \multicolumn{2}{c}{(0.1852)} && \multicolumn{2}{c}{(0.1054)} & \multicolumn{2}{c}{(0.0754)} \\
    AIC & $-$266&3646 & $-$180&6738 && $-$229&3934 & $-$167&8602 && $-$168&2601 & $-$186&7284 && $-$1&5742 & 1&1700 \vspace{2mm}\\
    \multicolumn{20}{l}{\textit{Panel E: Gumbel copula}}  \\
    $\alpha$ & \multicolumn{2}{c}{\textbf{3.1259}} & \multicolumn{2}{c}{\textbf{2.6813}} && \multicolumn{2}{c}{\textbf{2.8816}} & \multicolumn{2}{c}{\textbf{2.5287}} && \multicolumn{2}{c}{\textbf{2.3988}} & \multicolumn{2}{c}{\textbf{2.3370}} && \multicolumn{2}{c}{1.1052} & \multicolumn{2}{c}{1.0308} \\
        & \multicolumn{2}{c}{(0.1782)} & \multicolumn{2}{c}{(0.1557)} && \multicolumn{2}{c}{(0.1653)} & \multicolumn{2}{c}{(0.1469)} && \multicolumn{2}{c}{(0.1376)} & \multicolumn{2}{c}{(0.1300)} && \multicolumn{2}{c}{(0.0571)} & \multicolumn{2}{c}{(0.0429)} \\
    AIC & $-$320&1256 & $-$236&9790 && $-$284&9830 & $-$215&0106 && $-$198&3376 & $-$220&7230 && $-$2&4816 & 1&4400 \vspace{2mm}\\
    \multicolumn{20}{l}{\textit{Panel F: Survival Gumbel copula}}  \\
    $\alpha$ & \multicolumn{2}{c}{\textbf{3.1886}} & \multicolumn{2}{c}{\textbf{2.7764}} && \multicolumn{2}{c}{\textbf{3.0310}} & \multicolumn{2}{c}{\textbf{2.5553}} && \multicolumn{2}{c}{\textbf{2.3983}} & \multicolumn{2}{c}{\textbf{2.2744}} && \multicolumn{2}{c}{1.0873} & \multicolumn{2}{c}{1.0318} \\
        & \multicolumn{2}{c}{(0.1813)} & \multicolumn{2}{c}{(0.1602)} && \multicolumn{2}{c}{(0.1753)} & \multicolumn{2}{c}{(0.1460)} && \multicolumn{2}{c}{(0.1372)} & \multicolumn{2}{c}{(0.1263)} && \multicolumn{2}{c}{(0.0484)} & \multicolumn{2}{c}{(0.0435)} \\
    AIC & $-$319&7171 & $-$261&3444 && $-$296&3012 & $-$228&4302 && $-$194&8384 & $-$210&7936 && $-$4&4653$^{\#}$ & 1&3891 \\
    \bottomrule
    \end{tabular}
  \begin{flushleft}
    \footnotesize
    \justifying Note: This table reports the estimated parameters and goodness-of-fit measures of six different single copula models for each pair of the futures-spot returns before and after the outbreak of the Russia-Ukraine conflict, where the standard errors of parameter estimates are presented in parentheses. $\rho$ and $\alpha$ denote the copula parameters, and Dof is the degrees-of-freedom parameter of the Student-t copula model. AIC represent the value of the Akaike information criterion, and Bold numbers refer to significance at the 5\% level. The superscript $\#$ indicates the smallest AIC.
  \end{flushleft}
  \label{Tab:Agro_Single_Copula_Estimation}%
\end{table}%

Furthermore, the values of the AIC are also presented in Table~\ref{Tab:Agro_Single_Copula_Estimation} to assess the goodness of fit of the different single copula models before and after the outbreak of the Russia-Ukraine conflict. For the futures-spot pair of soybean, the Student-t copula yields the smallest AIC over the pre-outbreak period, followed by the Normal, Gumbel, and survival Gumbel copulas, which still performs best over the post-outbreak period, followed by the survival Gumbel, Normal, and Gumbel copulas. For the futures-spot pair of maize, the Student-t, survival Gumbel, and Gumbel copulas are always the three best-fitting models over both time periods. For the futures-spot pair of wheat, the Student-t copula results in the smallest AIC over the two periods, followed by the Normal, Gumbel, and survival Gumbel copulas. For the futures-spot pair of rice, the survival Gumbel, Student-t, and Gumbel copulas are the three best-fitting models over the pre-outbreak period, while the Student-t copula yields the smallest AIC over the post-outbreak period, followed by the survival Clayton, survival Gumbel, and Gumbel copulas.

However, the assumptions of tail independence in the Normal copula and symmetric tail dependence in the Student-t copula may be too restrictive for empirical analysis. Additionally, the Clayton and survival Gumbel copulas can only depict lower tail dependence, while the Gumbel and survival Clayton copulas can only capture upper tail dependence between variables. Therefore, to account for asymmetric tail dependence, we further construct mixed copula models to evaluate the upper and lower tail dependence between the agricultural futures and spot markets simultaneously.

\subsection{Estimation of mixed copula models}

On the basis of the analysis of single copula models, we further combine the Gumbel or survival Clayton copula, both of which are capable of describing upper tail dependence, with the Clayton or survival Gumbel copula, both of which can capture lower tail dependence. This allows us to construct different mixed copula models, thereby taking into account the possible asymmetric tail dependence between agricultural futures and spot returns.

Table~\ref{Tab:Agro_Mixed_Copula_Estimation} provides the estimated results of various mixed copula models for the futures-spot pairs of soybean, maize, wheat, and rice before and after the outbreak of the Russia-Ukraine conflict, respectively. As shown in Panel A, the mixed copula model combining the Gumbel and survival Gumbel copulas is the optimal mixed copula model with the smallest AIC for the futures-spot pairs of soybean, maize, wheat, and rice over the pre-outbreak period, as well as for the futures-spot pair of rice over the post-outbreak period. By contrast, the mixed copula model that combines the survival Clayton and survival Gumbel copulas yields the smallest AIC for the futures-spot pairs of soybean, maize, and wheat over the post-outbreak period, as indicated in Panel D.

In addition, the upper (lower) tail dependence refers to the probability that extreme rises (falls) in one market are accompanied by extreme rises (falls) in another market. Comparing the lower and upper tail dependence coefficients of the optimal mixed copula models for each pair over different periods, we note that the futures-spot pairs of the four agricultural commodities exhibit asymmetric tail dependence structures. Specifically, soybean futures and spots have a greater upper tail dependence over the pre-outbreak period, but a greater lower tail dependence over the post-outbreak period, similar to the tail dependence structure presented in wheat futures and spots. This implies that the interdependence between soybean futures and spots is more sensitive to bullish markets before the outbreak of the conflict, whereas it is more sensitive to bearish markets afterward. In contrast to soybean and wheat, the lower tail dependence between maize futures and spots is consistently larger than the upper tail dependence over the two sub-periods, indicating that maize markets tend to be more vulnerable to bad news than to good news. For rice futures and spots, the lower tail dependence is larger than the upper tail dependence before the outbreak of the conflict, whereas there is mainly an upper tail dependence and almost no lower tail dependence over the post-outbreak period.

\begin{table}[!ht]
  \centering
  \setlength{\abovecaptionskip}{0pt}
  \setlength{\belowcaptionskip}{10pt}
  \caption{Estimation of mixed copula models between futures and spot return series of agricultural commodities before and after the outbreak of the Russia-Ukraine conflict}
    \setlength\tabcolsep{3.8pt}
    \begin{tabular}{l r@{.}l r@{.}l c r@{.}l r@{.}l c r@{.}l  r@{.}l c r@{.}l r@{.}l}
    \toprule
        & \multicolumn{4}{c}{Soybean} && \multicolumn{4}{c}{Maize} && \multicolumn{4}{c}{Wheat} && \multicolumn{4}{c}{Rice} \\
        \cline{2-5}\cline{7-10}\cline{12-15}\cline{17-20}
        & \multicolumn{2}{c}{Before} & \multicolumn{2}{c}{After} && \multicolumn{2}{c}{Before} & \multicolumn{2}{c}{After} && \multicolumn{2}{c}{Before} & \multicolumn{2}{c}{After} && \multicolumn{2}{c}{Before} & \multicolumn{2}{c}{After}  \\
    \midrule
    \multicolumn{20}{l}{\textit{Panel A: Combining Gumbel copula and survival Gumbel copula}} \\
    $\theta_{c}^{1}$ & 3&5740 & 2&2059 && 4&1724 & 2&9213 && 2&5016 & 2&2698 && 1&0322 & 7&0818 \\
    $\theta_{c}^{2}$ & 3&2474 & 3&2456 && 2&7312 & 2&7019 && 2&5046 & 2&6949 && 1&6964 & 1&0000 \\
    $\omega_{1}$ & 0&5136 & 0&1815 && 0&4158 & 0&3863 && 0&5426 & 0&5326 && 0&8218 & 0&0577 \\
    $\omega_{2}$ & 0&4864 & 0&8185 && 0&5842 & 0&6137 && 0&4574 & 0&4674 && 0&1782 & 0&9423 \\
    $\lambda^{\mathrm{up}}$ & 0&4037 & 0&1145 && 0&3406 & 0&2829 && 0&3694 & 0&3424 && 0&0352 & 0&0517 \\
    $\lambda^{\mathrm{low}}$ & 0&3707 & 0&6237 && 0&4155 & 0&4342 && 0&3116 & 0&3303 && 0&0883 & 0&0000 \\
    $\tau$ & 0&7048 & 0&6632 && 0&6825 & 0&6391 && 0&5991 & 0&5901 && 0&0953 & 0&0381 \\
    AIC & $-$333&7438$^{\#}$ & $-$272&6353 && $-$302&0674$^{\#}$ & $-$236&9446 && $-$199&5487$^{\#}$ & $-$226&1020 && 0&5238$^{\#}$ & 4&0695$^{\#}$ \vspace{2mm} \\
    \multicolumn{20}{l}{\textit{Panel B: Combining survival Clayton copula and Clayton copula}} \\
    $\theta_{c}^{1}$ & 4&3818 & 4&5712 && 5&3217 & 4&0384 && 2&3631 & 2&4362 && 0&1193 & 5&6031 \\
    $\theta_{c}^{2}$ & 3&7329 & 3&4479 && 3&1603 & 2&9881 && 2&9257 & 2&6488 && 0&9533 & 0&0000 \\
    $\omega_{1}$ & 0&5146 & 0&2894 && 0&3847 & 0&4031 && 0&5471 & 0&5272 && 0&8407 & 0&0602 \\
    $\omega_{2}$ & 0&4854 & 0&7106 && 0&6153 & 0&5969 && 0&4529 & 0&4728 && 0&1593 & 0&9398 \\
    $\lambda^{\mathrm{up}}$ & 0&4393 & 0&2487 && 0&3377 & 0&3395 && 0&4080 & 0&3966 && 0&0025 & 0&0532 \\
    $\lambda^{\mathrm{low}}$ & 0&4031 & 0&5812 && 0&4942 & 0&4734 && 0&3573 & 0&3639 && 0&0770 & 0&0000 \\
    $\tau$ & 0&6619 & 0&6447 && 0&6480 & 0&6203 && 0&5602 & 0&5542 && 0&0975 & 0&0367 \\
    AIC & $-$315&1917 & $-$270&1799 && $-$280&1597 & $-$235&6581 && $-$186&8934 & $-$214&8656 && 0&8656 & 5&0964 \vspace{2mm} \\
    \multicolumn{20}{l}{\textit{Panel C: Combining Gumbel copula and Clayton copula}} \\
    $\theta_{c}^{1}$ & 3&5862 & 2&9692 && 3&0847 & 2&9800 && 2&4593 & 2&3769 && 1&0557 & 1&0308 \\
    $\theta_{c}^{2}$ & 3&5320 & 3&7904 && 4&0595 & 2&9585 && 3&1188 & 2&8349 && 0&9829 & 0&0153 \\
    $\omega_{1}$ & 0&7348 & 0&4667 && 0&6740 & 0&5654 && 0&7604 & 0&7690 && 0&8482 & 1&0000 \\
    $\omega_{2}$ & 0&2652 & 0&5333 && 0&3260 & 0&4346 && 0&2396 & 0&2310 && 0&1518 & 0&0000 \\
    $\lambda^{\mathrm{up}}$ & 0&5781 & 0&3440 && 0&5042 & 0&4173 && 0&5128 & 0&5086 && 0&0609 & 0&0410 \\
    $\lambda^{\mathrm{low}}$ & 0&2179 & 0&4441 && 0&2748 & 0&3438 && 0&1919 & 0&1809 && 0&0750 & 0&0000 \\
    $\tau$ & 0&6957 & 0&6548 && 0&6704 & 0&6312 && 0&5948 & 0&5788 && 0&0934 & 0&0299 \\
    AIC & $-$329&0310 & $-$270&4221 && $-$293&1911 & $-$235&7890 && $-$198&1150 & $-$220&1891 && 0&9953 & 7&4400 \vspace{2mm} \\
    \multicolumn{20}{l}{\textit{Panel D: Combining survival Clayton copula and survival Gumbel copula}} \\
    $\theta_{c}^{1}$ & 4&5883 & 4&4574 && 9&7557 & 4&1952 && 2&4501 & 2&3194 && 0&3572 & 5&6032 \\
    $\theta_{c}^{2}$ & 3&3621 & 3&0509 && 2&9227 & 2&7182 && 2&5279 & 2&6241 && 1&0872 & 1&0000 \\
    $\omega_{1}$ & 0&2594 & 0&1526 && 0&1870 & 0&2531 && 0&2822 & 0&3406 && 0&2875 & 0&0602 \\
    $\omega_{2}$ & 0&7406 & 0&8474 && 0&8130 & 0&7469 && 0&7178 & 0&6594 && 0&7125 & 0&9398 \\
    $\lambda^{\mathrm{up}}$ & 0&2230 & 0&1306 && 0&1742 & 0&2145 && 0&2127 & 0&2526 && 0&0413 & 0&0532 \\
    $\lambda^{\mathrm{low}}$ & 0&5711 & 0&6313 && 0&5954 & 0&5300 && 0&4913 & 0&4601 && 0&0770 & 0&0000 \\
    $\tau$ & 0&6978 & 0&6728 && 0&6839 & 0&6403 && 0&5868 & 0&5881 && 0&1008 & 0&0367 \\
    AIC & $-$330&8275 & $-$275&2252$^{\#}$ && $-$301&1977 & $-$239&7129$^{\#}$ && $-$194&2952 & $-$227&5526$^{\#}$ && 0&7220 & 5&0964 \\
    \bottomrule
    \end{tabular}%
  \begin{flushleft}
    \footnotesize
    \justifying Note: This table presents the parameter estimates and goodness-of-fit measures of four different mixed copula models for soybean, maize, wheat, and rice before and after the outbreak of the Russia-Ukraine conflict. $\theta_{c}^{1}$, $\theta_{c}^{2}$, $\omega_{1}$, and $\omega_{2}$ represent the copula parameters and weight parameters for the first and second component copulas, respectively. The upper tail dependence $\lambda^{\mathrm{up}}$ and the lower tail dependence $\lambda^{\mathrm{low}}$ for each agricultural pair are reported in the table, as well as the Kendall rank correlation coefficient $\tau$ and the Akaike information criterion AIC. The superscript $\#$ indicates the smallest AIC.
  \end{flushleft}
  \label{Tab:Agro_Mixed_Copula_Estimation}%
\end{table}%

\subsection{Tail dependence structure}

To incorporate both single copula and mixed copula models into our analytical framework, we compare the estimated results in Table~\ref{Tab:Agro_Single_Copula_Estimation} with those in Table~\ref{Tab:Agro_Mixed_Copula_Estimation} and find that, according to the Akaike information criterion, the futures-spot pairs of the four agricultural commodities before and after the outbreak of the Russia-Ukraine conflict correspond to different optimal copula models.

Specifically, the optimal model for soybeans over the pre-outbreak period is the Student-t copula model, while the mixed copula model combining the survival Clayton and survival Gumbel copulas is optimal over the post-outbreak period. For maize and wheat, the Student-t copula model consistently performs best before and after the outbreak of the conflict. For rice, the survival Gumbel copula model is the optimal model with the smallest AIC over the pre-outbreak period, but the Student-t copula model performs best over the post-outbreak period. In other words, soybean futures and spots exhibit symmetric tail dependence before the outbreak of the conflict, but asymmetric tail dependence with lower tail dependence greater than upper tail dependence afterward. Maize and wheat consistently display symmetric tail dependence over the pre- and post-outbreak periods. Rice futures and spots show lower tail dependence and symmetric tail dependence before and after the outbreak of the conflict, respectively.


\subsection{Risk spillover measurement}

Based on the estimated results of the optimal copula models corresponding to soybean, maize, wheat, and rice over the pre- and post-outbreak periods, we further calculate the risk measures $VaR$, $CoVaR$, and $\Delta CoVaR$ for each agricultural spot market to quantify the extreme downside and upside risk spillover effects of the futures market on the spot market before and after the outbreak of the Russia-Ukraine conflict, respectively.

\begin{figure}[!t]
\centering
\includegraphics[width=0.475\linewidth]{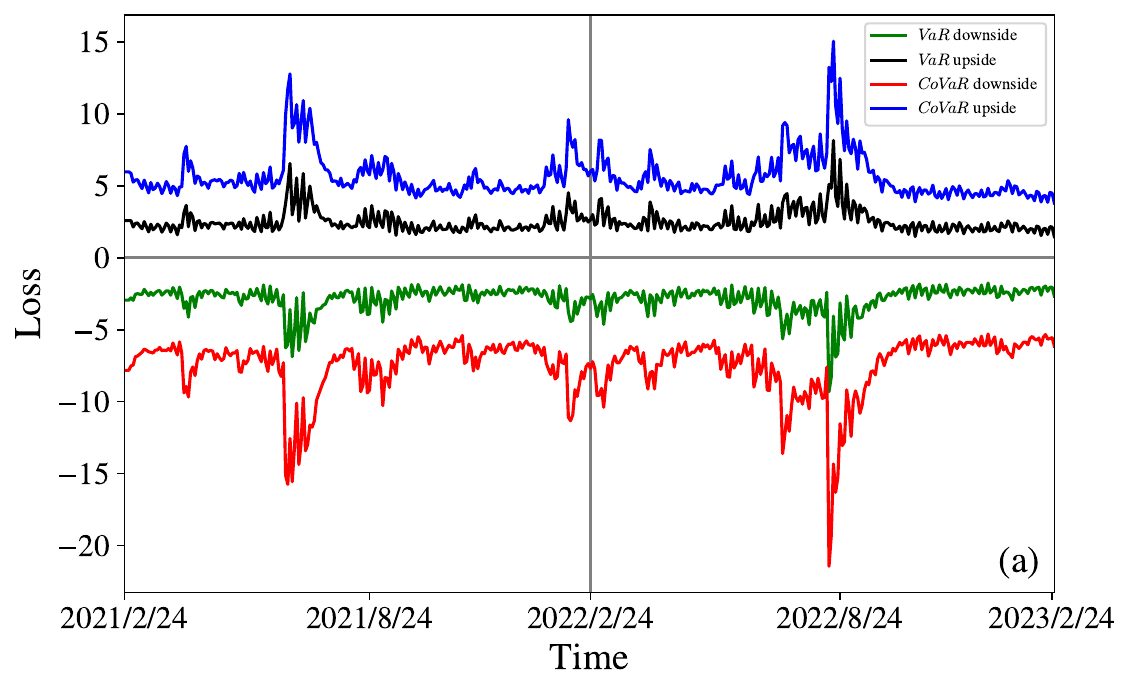}
\includegraphics[width=0.475\linewidth]{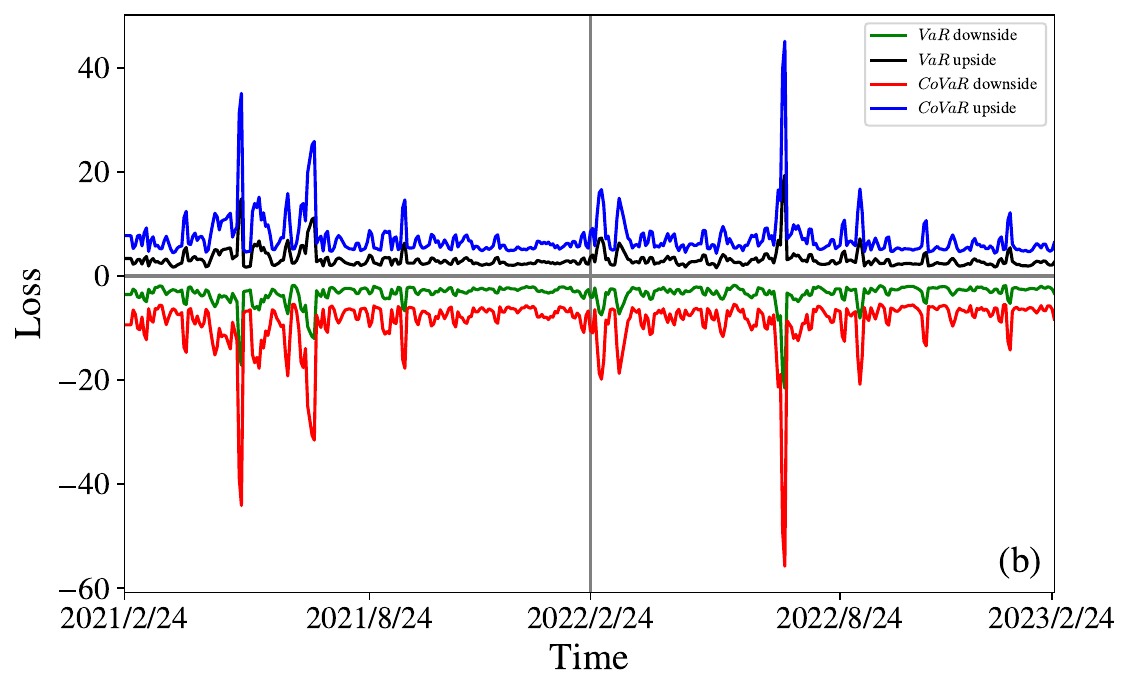}
\includegraphics[width=0.475\linewidth]{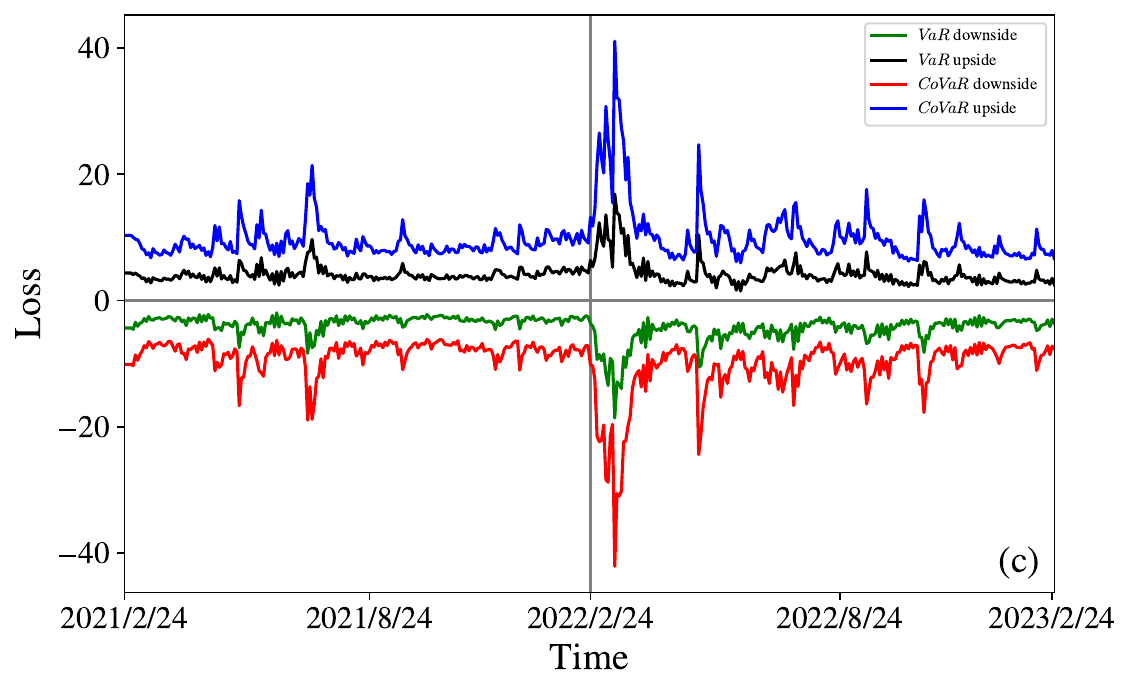}
\includegraphics[width=0.475\linewidth]{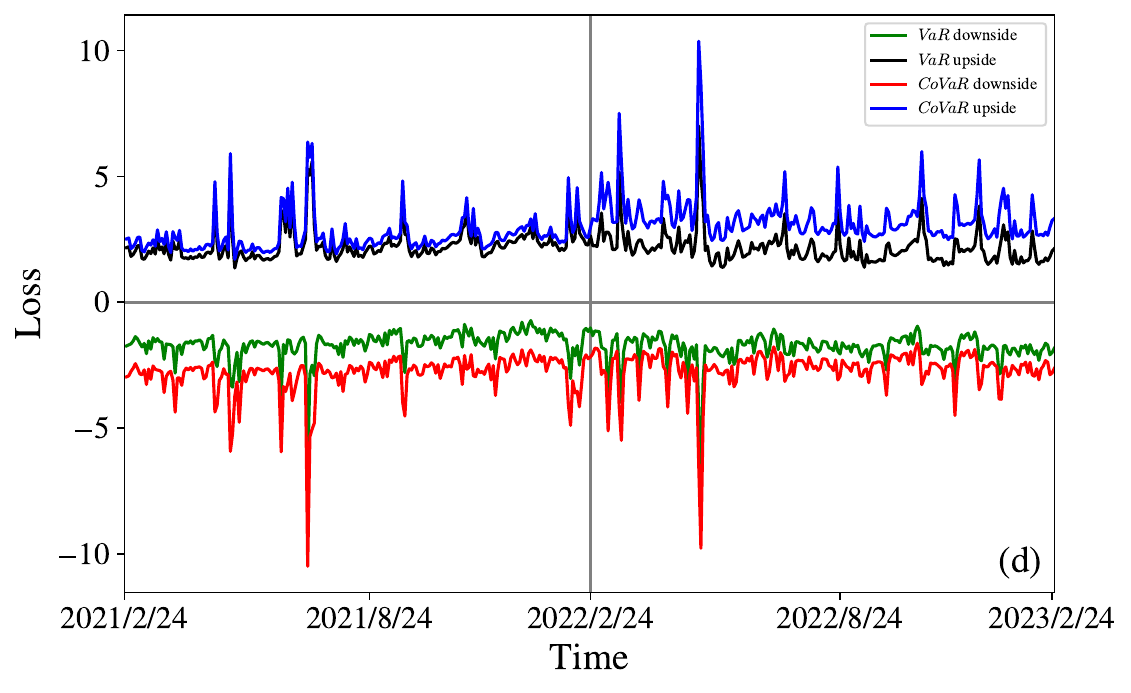}
\caption{Dynamics of the downside and upside $VaR$ and $CoVaR$ for soybean (a), maize (b), wheat (c), and rice (d) before and after the outbreak of the Russia-Ukraine conflict, where the vertical line indicates the date of the outbreak of the conflict.}
\label{Fig:Agro_Risk_Measures}
\end{figure}

Figure~\ref{Fig:Agro_Risk_Measures} illustrates the dynamics of $VaR$ and $CoVaR$ for soybean, maize, wheat, and rice spots from February 24, 2021, to February 24, 2023. A comparison of the results reveals that, although the ranges of $VaR$ and $CoVaR$ vary across agricultural commodities, the downside $CoVaR$s are consistently smaller than the corresponding downside $VaR$s for each commodity, while the upside $CoVaR$s are always larger than the corresponding upside $VaR$s. This finding implies that extreme downside and upside movements in agricultural futures prices exert an influence on agricultural spot prices, suggesting extreme risk spillovers from the agricultural futures market to the corresponding spot market. Moreover, we observe that the downside $VaR$s and $CoVaR$s for all agricultural commodities decrease, while their upside $VaR$s and $CoVaR$s increase after the outbreak of the Russia-Ukraine conflict, especially for wheat, whose downside $VaR$s, $CoVaR$s, and upside $VaR$s and $CoVaR$s reach their troughs and peaks within a short period, respectively. In other words, the outbreak of the conflict has exacerbated risks in international markets for soybean, maize, wheat, and rice to varying degrees, with the wheat market being hit the hardest. For one thing, Russia, the world's largest wheat exporter, and Ukraine, known as the breadbasket of Europe, are major exporters of wheat, jointly accounting for approximately one-third of global wheat output. However, the Russia-Ukraine conflict has caused serious damage to grain production and exports in both countries. For another, the conflict has led to a sharp rise in energy prices and, thus, increased demand for biomass fuels made from wheat and other grains, which further aggravates the tension between global wheat supply and demand. Consequently, the outbreak of the Russia-Ukraine conflict has severely impacted the global wheat market, resulting in a remarkable increase in the risks of the international wheat market.

In addition, compared with the pre-outbreak period, the differences between the downside $VaR$s and $CoVaR$s for rice spots have significantly decreased over the post-outbreak period, while the differences between the upside $VaR$s and $CoVaR$s have markedly increased. These changes reflect the reaction of the rice futures and spot markets to the Russia-Ukraine conflict, including the weakening of the downside risk spillover effect and the strengthening of the upside risk spillover effect. This finding is consistent with the graphical evidence shown in Figure~\ref{Fig:AgroPrice_evolution}(d), where the prices of rice futures and spots have been on an upward oscillating trend since the outbreak of the Russia-Ukraine conflict. Actually, the rise in international rice prices is driven not only by the horizontal price transmission of wheat and other related varieties, but also by market forces consisting of production cost push and demand pull. On the production side, the main rice-producing countries are usually fertilizer importers, who are under enormous pressure from agricultural costs. The conflict has blocked the exports of major fertilizer-producing countries such as Russia and Belarus, thereby pushing up the prices of agricultural materials and boosting the cost of rice production. On the demand side, considering the rising prices of wheat and maize, some countries have begun to use rice as an alternative, resulting in increasing market demand for rice. However, as Figure~\ref{Fig:Agro_Risk_Measures}(d) shows, the values of $VaR$s and $CoVaR$s for rice are small with limited variation ranges, which can be largely attributed to the relative independence of the rice market from other agricultural markets. Rice is mostly consumed as a grain ration with a comparatively short processing chain, and its total demand remains stable due to dietary habits. Moreover, given constraints such as nutritive indices and energy values, rice does not have the potential to be widely used as feed. Rice production and consumption are dominated by Asian countries, most of which basically ensure domestic self-sufficiency. As a result, the international rice market is less risky than the soybean, maize, and wheat markets.

Further analysis of Figure~\ref{Fig:Agro_Risk_Measures} shows that besides the heightened volatility around the outbreak of the Russia-Ukraine conflict, the $VaR$s and $CoVaR$s for the spot returns of soybean, maize, wheat, and rice experienced significant fluctuations from May to July 2021, which can be mainly attributed to the combined effects of weather extremes and the COVID-19 pandemic. Global warming and the prolonged La Niña phenomenon have led to frequent extreme weather events worldwide, especially since June 2021, including continuous torrential rain and serious flooding in India, Germany, and China, record-breaking high temperatures and droughts in the United States and Turkey, and massive wildfires in Greece, Brazil, the United States, and Argentina. Droughts, floods, and other disasters drastically reduced grain yields in the world's main producing areas, triggering a tightening of international food supplies. {\textit{The State of Food Security and Nutrition in the World 2022}}\footnote{\url{https://www.fao.org/publications/sofi/en/}}, jointly released by several UN agencies such as the FAO and WFP, also points out that the global food system suffered a heavy blow from frequent weather extremes in 2021. Moreover, disruptions in the food supply chain owing to the COVID-19 pandemic further widened the food supply-demand gap, contributed to high and volatile food prices, and thus exacerbated the risks in global food markets.

As Figure~\ref{Fig:Agro_Risk_Measures}(a) shows, in August 2022, the $VaR$s and $CoVaR$s for soybean spot returns exhibited a significant increase in volatility and attained their respective extreme values, which can be explained by the compounding effects of multiple factors. Continuous drought and flooding in early 2022 considerably reduced soybean harvests in South America, particularly in Brazil and Argentina, resulting in a decline in global soybean output. Furthermore, the unstable expectations of soybean production, transmission of fluctuations from energy and financial markets, and high prices of fertilizers resulting from the Russia-Ukraine conflict jointly intensified the volatility of the soybean market. Additionally, the hot and dry weather in the United States soybean-producing regions in August 2022 further provoked the heating-up of speculative activity, aggravating the risks of the international soybean market.

As can be seen from Figure~\ref{Fig:Agro_Risk_Measures}(b), the downside $VaR$s, $CoVaR$s, and upside $VaR$s and $CoVaR$s for maize spot returns experienced sharp changes, reaching their lowest and highest points in July 2022, respectively. These fluctuations can be primarily attributed to extreme weather events and the evolution of the Russia-Ukraine conflict. On the one hand, the 2022 spring crop sowing area in Ukraine failed to meet expectations because of the conflict, and in July 2022, extreme heat and dry weather in the European Union and the United States, coupled with increased planting costs, led to a drastic decline in maize production and export volumes. Moreover, given that maize is a substitute feed grain for wheat, high international wheat prices indirectly drove market demand for maize. Further tightening of the global maize supply and demand contributed to a rapid increase in the risks of the international maize market. On the other hand, through active mediation by the United Nations and Turkey, Russia and Ukraine held four-party talks on July 13, 2022, and signed the Black Sea Grain Initiative on July 22, 2022, aiming to resume unimpeded exports of food and fertilizer from Black Sea ports. With the promotion of the Black Sea Grain Initiative, 38.94 million tons of agricultural products were shipped through the Black Sea corridor from February to December 2022, including 15.6 million tons of maize, 8.6 million tons of wheat, and 1.7 million tons of soybeans. In a sense, this initiative has brought relief and stability to global food markets, and has played a positive role in alleviating global food shortages and addressing the global food crisis.

Turning to Figure~\ref{Fig:Agro_Risk_Measures}(c), it is evident that the $VaR$s and $CoVaR$s for wheat spot returns fluctuated markedly in May 2022, which can be considered as a result of multiple uncertainty factors, such as climate change and the Russia-Ukraine conflict. Frequent weather extremes and the escalating conflict have resulted in a resurgence of food trade protectionism. According to the International Food Policy Research Institute (IFPRI), as of May 28, 2022, more than 20 countries, such as Argentina, Kazakhstan, India, Indonesia, and Turkey, have implemented export restrictions on food, including wheat, maize, flour, and beans. India, the world's second-largest wheat producer, announced a ban on wheat exports on May 13, 2022, owing to the impact of exceptional heat, worsening the global shortage of wheat supply. Moreover, several major wheat-producing regions such as the United States and France suffered from high temperatures and droughts, leading to lower wheat production. Consequently, the gap between global wheat supply and demand further widened, exacerbating the risks of the international wheat market.

As shown in Figure~\ref{Fig:Agro_Risk_Measures}(d), the downside $VaR$s, $CoVaR$s, and upside $VaR$s and $CoVaR$s for rice spot returns were highly volatile in May 2022, and even reached their respective troughs and peaks, the reason for which lies in rising food trade protectionism caused by extreme weather and the ongoing Russia-Ukraine conflict. India, Vietnam, and Thailand are the top three rice exporters in the world, accounting for 70\% of global rice exports in 2021. On May 27, 2022, the Thai and Vietnamese governments announced a joint plan to raise rice prices in the domestic market. To cope with the decrease in rice production due to insufficient rainfall and severe drought, as well as to ease domestic inflationary pressure, India also claimed that it was considering restricting rice exports in May 2022, and officially announced a ban on the export of broken rice, along with a 20\% tariff imposed on exports of other types of rice, on September 9, 2022. In June 2022, some major rice-producing countries, such as China and Pakistan, were hit by heavy rains and floods, resulting in potential yield losses for rice. As a result, global rice production and stocks decreased, while consumption and demand continued to increase, partly intensifying the risks of the international rice market.

Table~\ref{Tab:Statistics_VaR} reports the descriptive statistics of $VaR$, $CoVaR$, and $\Delta CoVaR$ for spot returns of soybean, maize, wheat, and rice before and after the outbreak of the Russia-Ukraine conflict, including the maximum, minimum, mean, and standard deviation. We observe that, for each agricultural commodity, the maximum, minimum, and mean values of its downside $CoVaR$s are smaller than the corresponding values of its downside $VaR$s over each sub-period, whereas the maximum, minimum, and mean values of its upside $CoVaR$s are larger than the corresponding values of its upside $VaR$s. Meanwhile, the maximum value of its downside $\Delta CoVaR$s is less than 0, and the minimum value of its upside $\Delta CoVaR$s is greater than 0. These results reaffirm the graphical evidence shown in Figure~\ref{Fig:Agro_Risk_Measures} that there exist extreme downside and upside risk spillover effects from agricultural futures markets to their corresponding agricultural spot markets over both sub-periods. It is also noteworthy that the summary statistics of $VaR$, $CoVaR$, and $\Delta CoVaR$ for rice spots are small compared to soybean, maize, and wheat, implying relatively limited risk in the international rice market. Furthermore, except for the downside $CoVaR$s and $\Delta CoVaR$s for rice spots, the standard deviations of the $VaR$s, $CoVaR$s, and $\Delta CoVaR$s for the other series over the post-outbreak period are larger than those of their corresponding $VaR$s, $CoVaR$s, and $\Delta CoVaR$s over the pre-outbreak period, which indicates that the changes in risks in the global food market are more pronounced after the outbreak of the Russia-Ukraine conflict.

\begin{table}[!ht]
  \centering
  \setlength{\abovecaptionskip}{0pt}
  \setlength{\belowcaptionskip}{10pt}
  \caption{Summary statistics of the risk measures $VaR$, $CoVaR$ and $\Delta CoVaR$ before and after the outbreak of the Russia-Ukraine conflict}
  \setlength\tabcolsep{5.8pt}
    \begin{tabular}{l r@{.}l r@{.}l c r@{.}l r@{.}l c r@{.}l  r@{.}l c r@{.}l r@{.}l}
    \toprule
        & \multicolumn{4}{c}{Max} && \multicolumn{4}{c}{Min} && \multicolumn{4}{c}{Mean} && \multicolumn{4}{c}{Std. Dev.} \\
        \cline{2-5}\cline{7-10}\cline{12-15}\cline{17-20}
        & \multicolumn{2}{c}{Before} & \multicolumn{2}{c}{After} && \multicolumn{2}{c}{Before} & \multicolumn{2}{c}{After} && \multicolumn{2}{c}{Before} & \multicolumn{2}{c}{After} && \multicolumn{2}{c}{Before} & \multicolumn{2}{c}{After} \\
    \midrule
    \multicolumn{20}{l}{\textit{Panel A: Soybean}}  \\
    $VaR_{0.05}$ & $-$1&8548 & $-$1&7530 && $-$6&8759 & $-$9&3114 && $-$2&8019 & $-$2&9162 && 0&8203 & 1&0306 \\
    $VaR_{0.95}$ & 6&5630 & 8&1618 && 1&5590 & 1&4749 && 2&4683 & 2&5509 && 0&7161 & 0&8685 \\
    $CoVaR_{0.05}$ & $-$5&3989 & $-$5&2964 && $-$15&7514 & $-$21&4473 && $-$7&4500 & $-$7&5176 && 1&8513 & 2&3063 \\
    $CoVaR_{0.95}$ & 12&7878 & 15&0584 && 4&1562 & 3&8113 && 5&6963 & 5&6380 && 1&4035 & 1&6909 \\
    $\Delta CoVaR_{0.05}$ & $-$2&8555 & $-$2&6619 && $-$7&8600 & $-$9&7018 && $-$3&7579 & $-$3&6785 && 0&8947 & 1&0806 \\
    $\Delta CoVaR_{0.95}$ & 5&4367 & 6&4292 && 1&9751 & 1&7640 && 2&5993 & 2&4377 && 0&6189 & 0&7161 \vspace{1mm}\\
    \multicolumn{20}{l}{\textit{Panel B: Maize}}  \\
    $VaR_{0.05}$ & $-$1&8179 & $-$1&7668 && $-$17&1451 & $-$21&5506 && $-$3&4881 & $-$3&2761 && 1&9497 & 2&0518 \\
    $VaR_{0.95}$ & 14&8389 & 19&3788 && 1&6862 & 1&5761 && 3&2723 & 3&1214 && 1&7635 & 1&8185 \\
    $CoVaR_{0.05}$ &$-$5&4457 & $-$5&4041 && $-$44&0867 & $-$55&7593 && $-$9&1826 & $-$8&6231 && 5&0384 & 5&2445 \\
    $CoVaR_{0.95}$ & 35&0943 & 45&0994 && 4&4446 & 4&3489 && 7&5536 & 7&1417 && 4&0830 & 4&2157 \\
    $\Delta CoVaR_{0.05}$ & $-$2&8995 & $-$2&8671 && $-$21&8084 & $-$27&6544 && $-$4&6096 & $-$4&3225 && 2&5084 & 2&5897 \\
    $\Delta CoVaR_{0.95}$ & 16&3449 & 20&7277 && 2&1731 & 2&1489 && 3&4548 & 3&2399 && 1&8800 & 1&9410 \vspace{1mm}\\
    \multicolumn{20}{l}{\textit{Panel C: Wheat}}  \\
    $VaR_{0.05}$ & $-$1&9362 & $-$2&3706 && $-$8&3522 & $-$18&6023 && $-$3&3060 & $-$4&8833 && 0&9159 & 2&2734 \\
    $VaR_{0.95}$ & 9&6991 & 16&8855 && 2&4397 & 1&5106 && 4&1079 & 4&3016 && 0&9285 & 2&2643 \\
    $CoVaR_{0.05}$ & $-$6&0940 & $-$6&5653 && $-$18&9499 & $-$42&0764 && $-$8&2231 & $-$10&9588 && 1&9255 & 5&0665 \\
    $CoVaR_{0.95}$ & 21&4182 & 41&0782 && 6&7706 & 5&9702 && 9&1756 & 10&5631 && 1&9714 & 5&1436 \\
    $\Delta CoVaR_{0.05}$ & $-$3&0481 & $-$2&9959 && $-$9&0478 & $-$18&7217 && $-$3&9125 & $-$4&8455 && 0&8533 & 2&2770 \\
    $\Delta CoVaR_{0.95}$ & 9&3281 & 19&3016 && 3&1425 & 3&0887 && 4&0338 & 4&9956 && 0&8797 & 2&3475 \vspace{1mm}\\
    \multicolumn{20}{l}{\textit{Panel D: Rice}}  \\
    $VaR_{0.05}$ & $-$0&7243 & $-$0&9483 && $-$6&7083 & $-$7&2823 && $-$1&6355 & $-$1&8279 && 0&5708 & 0&6011 \\
    $VaR_{0.95}$ & 5&5906 & 7&0063 && 1&3609 & 1&3835 && 2&2912 & 2&1594 && 0&6517 & 0&6586 \\
    $CoVaR_{0.05}$ & $-$1&8855 & $-$1&6269 && $-$10&4867 & $-$9&7726 && $-$2&8715 & $-$2&6372 && 0&8452 & 0&7870 \\
    $CoVaR_{0.95}$ & 6&3698 & 10&3708 && 1&7074 & 2&4150 && 2&6428 & 3&3619 && 0&7297 & 0&9455 \\
    $\Delta CoVaR_{0.05}$ & $-$0&9018 & $-$0&6248 && $-$3&2547 & $-$2&3042 && $-$1&0647 & $-$0&7488 && 0&2730 & 0&1985 \\
    $\Delta CoVaR_{0.95}$ & 0&6470 & 3&4271 && 0&1793 & 0&9292 && 0&2116 & 1&1137 && 0&0543 & 0&2952 \\
    \bottomrule
    \end{tabular}
  \begin{flushleft}
    \footnotesize
    \justifying Note: This table provides the summary statistics of the downside and upside $VaR$, $CoVaR$ and $\Delta CoVaR$ for the spot returns of soybean, maize, wheat, and rice before and after the outbreak of the Russia-Ukraine conflict, based on their corresponding optimal copula models. The subscripts 0.05 and 0.95 denote the downside and upside risk measures, respectively.
  \end{flushleft}
  \label{Tab:Statistics_VaR}
\end{table}

Having confirmed the existence of extreme risk spillovers from agricultural futures returns to spot returns, we further apply the K-S test to assess the significance of these risk spillover effects, along with the possible asymmetries across them at both the directional (downside versus upside) and temporal (pre-outbreak versus post-outbreak) levels. The hypothesis testing results are shown in Table~\ref{Tab:Spillover_Test}. Panels A and B display the test results for the significance of downside and upside risk spillover effects in the soybean, maize, wheat, and rice markets before and after the outbreak of the Russia-Ukraine conflict, respectively. As can be seen, all the K-S statistics are significant at the 1\% level, meaning that for each agricultural commodity, its futures market exhibits significant extreme downside and upside risk spillover effects on the corresponding spot market for both the pre- and post-outbreak periods. To put it another way, information from agricultural futures markets tends to significantly increase the risk exposure of the corresponding agricultural spot markets, given the co-movements of both markets.

\begin{table}[!ht]
  \centering
  \setlength{\abovecaptionskip}{0pt}
  \setlength{\belowcaptionskip}{10pt}
  \caption{Hypothesis testing for downside and upside risk spillover effects before and after the outbreak of the Russia-Ukraine conflict}
    \setlength\tabcolsep{12pt}
    \begin{tabular}{l r@{.}l r@{.}l r@{.}l r@{.}l}
    \toprule
        & \multicolumn{2}{c}{Soybean} & \multicolumn{2}{c}{Maize} & \multicolumn{2}{c}{Wheat} & \multicolumn{2}{c}{Rice} \\
    \midrule
    \multicolumn{9}{l}{\textit{Panel A: Tests for risk spillover effects in the pre-outbreak period}} \vspace{1mm} \\
    \makecell[l]{$H_{0}: CoVaR_{0.05} \geq VaR_{0.05}$ \\ $H_{1}: CoVaR_{0.05} < VaR_{0.05}$} & 0&9763$^{***}$ & 0&9100$^{***}$ & 0&9810$^{***}$ & 0&8863$^{***}$ \vspace{1mm} \\
    \makecell[l]{$H_{0}: CoVaR_{0.95} \leq VaR_{0.95}$ \\ $H_{1}: CoVaR_{0.95} > VaR_{0.95}$} & 0&9716$^{***}$ & 0&8673$^{***}$ & 0&9810$^{***}$ & 0&3460$^{***}$ \vspace{2mm}\\
    \multicolumn{9}{l}{\textit{Panel B: Tests for risk spillover effects in the post-outbreak period}} \vspace{1mm} \\
    \makecell[l]{$H_{0}: CoVaR_{0.05} \geq VaR_{0.05}$ \\ $H_{1}: CoVaR_{0.05} < VaR_{0.05}$} & 0&9619$^{***}$ & 0&9381$^{***}$ & 0&8714$^{***}$ & 0&7143$^{***}$ \vspace{1mm} \\
    \makecell[l]{$H_{0}: CoVaR_{0.95} \leq VaR_{0.95}$ \\ $H_{1}: CoVaR_{0.95} > VaR_{0.95}$} & 0&9238$^{***}$ & 0&9095$^{***}$ & 0&8952$^{***}$ & 0&8238$^{***}$ \vspace{2mm} \\
    \multicolumn{9}{l}{\textit{Panel C: Tests for downside and upside risk spillover effects in the pre-outbreak period}} \vspace{1mm} \\
    \makecell[l]{$H_{0}: {CoVaR_{0.05}} \Big/ {VaR_{0.05}} \leq {CoVaR_{0.95}} \Big/ {VaR_{0.95}}$ \\ $H_{1}: {CoVaR_{0.05}} \Big/ {VaR_{0.05}} > {CoVaR_{0.95}} \Big/ {VaR_{0.95}}$} & 0&6683$^{***}$ & 0&8341$^{***}$ & 0&6209$^{***}$ & 1&0000$^{***}$ \vspace{1mm} \\
    \makecell[l]{$H_{0}: {CoVaR_{0.05}} \Big/ {VaR_{0.05}} \geq {CoVaR_{0.95}} \Big/ {VaR_{0.95}}$ \\ $H_{1}: {CoVaR_{0.05}} \Big/ {VaR_{0.05}} < {CoVaR_{0.95}} \Big/ {VaR_{0.95}}$} & 0&0000 & 0&0000 & 0&0000 & 0&0000 \vspace{2mm} \\
    \multicolumn{9}{l}{\textit{Panel D: Tests for downside and upside risk spillover effects in the post-outbreak period}} \vspace{1mm} \\
    \makecell[l]{$H_{0}: {CoVaR_{0.05}} \Big/ {VaR_{0.05}} \leq {CoVaR_{0.95}} \Big/ {VaR_{0.95}}$ \\ $H_{1}: {CoVaR_{0.05}} \Big/ {VaR_{0.05}} > {CoVaR_{0.95}} \Big/ {VaR_{0.95}}$} & 0&6143$^{***}$ & 0&8191$^{***}$ & 0&0000 & 0&0000 \vspace{1mm} \\
    \makecell[l]{$H_{0}: {CoVaR_{0.05}} \Big/ {VaR_{0.05}} \geq {CoVaR_{0.95}} \Big/ {VaR_{0.95}}$ \\ $H_{1}: {CoVaR_{0.05}} \Big/ {VaR_{0.05}} < {CoVaR_{0.95}} \Big/ {VaR_{0.95}}$} & 0&0000 & 0&0000 & 0&5143$^{***}$ & 0&5143$^{***}$ \vspace{2mm} \\
    \multicolumn{9}{l}{\textit{Panel E: Tests for downside risk spillover effects between the pre- and post-outbreak periods}} \vspace{1mm} \\
    \makecell[l]{$H_{0}: {CoVaR^{\text{pre}}_{0.05}} \Big/ {VaR^{\text{pre}}_{0.05}} \leq {CoVaR^{\text{post}}_{0.05}} \Big/ {VaR^{\text{post}}_{0.05}}$ \\ $H_{1}: {CoVaR^{\text{pre}}_{0.05}} \Big/ {VaR^{\text{pre}}_{0.05}} > {CoVaR^{\text{post}}_{0.05}} \Big/ {VaR^{\text{post}}_{0.05}}$} & 0&2152$^{***}$ & 0&0817 & 0&5534$^{***}$ & 0&8861$^{***}$ \vspace{1mm} \\
    \makecell[l]{$H_{0}: {CoVaR^{\text{pre}}_{0.05}} \Big/ {VaR^{\text{pre}}_{0.05}} \geq {CoVaR^{\text{post}}_{0.05}} \Big/ {VaR^{\text{post}}_{0.05}}$ \\ $H_{1}: {CoVaR^{\text{pre}}_{0.05}} \Big/ {VaR^{\text{pre}}_{0.05}} < {CoVaR^{\text{post}}_{0.05}} \Big/ {VaR^{\text{post}}_{0.05}}$} & 0&0048 & 0&0920 & 0&0000 & 0&0000 \vspace{2mm} \\
    \multicolumn{9}{l}{\textit{Panel F: Tests for upside risk spillover effects between the pre- and post-outbreak periods}} \vspace{1mm} \\
    \makecell[l]{$H_{0}: {CoVaR^{\text{pre}}_{0.95}} \Big/ {VaR^{\text{pre}}_{0.95}} \leq {CoVaR^{\text{post}}_{0.95}} \Big/ {VaR^{\text{post}}_{0.95}}$ \\ $H_{1}: {CoVaR^{\text{pre}}_{0.95}} \Big/ {VaR^{\text{pre}}_{0.95}} > {CoVaR^{\text{post}}_{0.95}} \Big/ {VaR^{\text{post}}_{0.95}}$} & 0&2625$^{***}$ & 0&1488$^{***}$ & 0&0000 & 0&0000 \vspace{1mm} \\
    \makecell[l]{$H_{0}: {CoVaR^{\text{pre}}_{0.95}} \Big/ {VaR^{\text{pre}}_{0.95}} \geq {CoVaR^{\text{post}}_{0.95}} \Big/ {VaR^{\text{post}}_{0.95}}$ \\ $H_{1}: {CoVaR^{\text{pre}}_{0.95}} \Big/ {VaR^{\text{pre}}_{0.95}} < {CoVaR^{\text{post}}_{0.95}} \Big/ {VaR^{\text{post}}_{0.95}}$} & 0&0000 & 0&0295 & 0&5486$^{***}$ & 1&0000$^{***}$ \\
    \bottomrule
    \end{tabular}
  \begin{flushleft}
    \footnotesize
    \justifying Note: This table presents the K-S test results for the downside and upside risk spillover effects before and after the outbreak of the Russia-Ukraine conflict. The K-S tests in Panels A and B explore the significance of the two effects over the pre- and post-outbreak periods, respectively. The K-S tests in Panels C and D assess the strength differences between the downside and upside risk spillover effects over the two periods, while the K-S tests in Panels E and F evaluate the strength differences of the two effects between the pre- and post-outbreak periods. The superscripts pre and post refer to the measures before and after the outbreak of the conflict, and the subscripts 0.05 and 0.95 indicate the downside and upside risk measures, respectively. $^{***}$ denotes significance at the 1\% level.
  \end{flushleft}
  \label{Tab:Spillover_Test}
\end{table}

Panel C explores the possible asymmetry between the downside and upside risk spillover effects for each agricultural commodity before the outbreak of the Russia-Ukraine conflict, while Panel D investigates the possible asymmetry between the two risk spillover effects after the outbreak. From the test results in Panel C, it can be found that the downside risk spillover effects are stronger than their corresponding upside risk spillover effects for the four agricultural commodities over the pre-outbreak period. As shown in Panel D, the downside risk spillover effects of soybean and maize remain stronger than their respective upside risk spillover effects after the outbreak of the conflict. However, for wheat and rice, the situation is reversed, with their upside risk spillover effects being stronger than the corresponding downside risk spillover effects. These test results suggest that the extreme downside risk spillovers in the soybean and maize markets remain dominant during both the pre- and post-outbreak periods, whereas in the wheat and rice markets, their extreme upside risk spillovers have relatively intensified since the outbreak of the Russia-Ukraine conflict. Given the stronger downside risk spillover effects between their futures and spots, heightened attention should always be paid to extreme downside risks in the soybean and maize markets. Additionally, extreme upside risks in the wheat and rice markets also deserve greater consideration, as their strength of upside risk spillovers exceeds that of downside risk spillovers after the outbreak of the conflict, which may be related to the surge in wheat prices and the oscillating rise in rice prices.

Panel E examines the strength difference in downside risk spillover effects between the pre- and post-outbreak periods for each agricultural commodity, while Panel F evaluates the strength difference in upside risk spillover effects before and after the outbreak of the Russia-Ukraine conflict. Concerning the extreme downside risk spillovers, there is basically no discernible difference in the strength of these effects in the maize market between the two periods, as shown in Panel E. However, in the soybean, wheat, and rice markets, the downside risk spillover effects before the outbreak of the conflict are more pronounced and robust than those observed afterward. Regarding the extreme upside risk spillovers analyzed in Panel F, the soybean and maize markets exhibit notably stronger upside risk spillover effects over the pre-outbreak period, whereas the wheat and rice markets, conversely, demonstrate significantly stronger upside risk spillover effects over the post-outbreak period. According to these findings, we can draw the conclusion that the international markets for the four major grain crops have all reacted to the outbreak of the Russia-Ukraine conflict in some manner. That is to say, the conflict between Russia and Ukraine has impacted the situation of extreme risk spillovers within these international agricultural markets to varying degrees. Specifically, following the outbreak of the conflict, the soybean market has witnessed a relative weakening of both downside and upside risk spillover effects, as well as a reduction in the strength of the upside risk spillover effect in the maize market. Furthermore, the downside risk spillover effects in the wheat and rice markets have relatively diminished, whereas their upside risk spillover effects have relatively strengthened, in line with the test results observed in Panels C and D.

\section{Conclusions}
\label{S1:Conclude}

Focusing on the impact of the Russia-Ukraine conflict on the global food market, we utilize the Copula-CoVaR approach to examine the tail dependence structures and extreme risk spillovers between futures and spots of four main grain crops (soybean, maize, wheat, and rice) over the pre- and post-outbreak periods. Considering data characteristics such as serial autocorrelation and conditional heteroskedasticity, we first employ the ARMA-GARCH-skewed Student-t model to specify the marginal distribution of each agricultural return series. To clarify the tail dependence between agricultural futures and spots more accurately and comprehensively, we propose a new analytical framework that incorporates various single and mixed copula models simultaneously, and determines the optimal copula model for each agricultural return series. Our empirical analysis shows that the futures-spot pairs of the four agricultural commodities over the two sub-periods correspond to different optimal copula models, implying that each staple crop exhibits a particular tail dependence structure. Specifically, soybean futures and spots show symmetric tail dependence over the pre-outbreak period, but asymmetric tail dependence with lower tail dependence being larger than upper tail dependence over the post-outbreak period. The futures and spots of maize and wheat consistently present symmetric tail dependence before and after the conflict outbreak. As for rice, its futures and spots display mainly lower tail dependence over the pre-outbreak period but symmetric tail dependence afterward.

With the parameter estimates of marginal and copula models, we further quantify the downside and upside risk measures, including $VaR$, $CoVaR$, and $\Delta CoVaR$, for the international soybean, maize, wheat, and rice markets over the pre- and post-outbreak periods. Based on the empirical results, we can safely conclude that the Russia-Ukraine conflict has exacerbated the market risks of these staple crops to varying degrees, with the wheat market being hit the hardest. In particular, the reaction of the international rice market to the conflict is reflected in the weakening of the downside risk spillover effect and the strengthening of the upside risk spillover effect, which are consistent with the upward oscillating trend of rice prices since the outbreak of the conflict. Additionally, combined with the dynamic evolution of risk measures for the four agricultural commodities, we provide some reasonable explanations for the noticeable increases in the risks of the soybean market in August 2022, the maize market in July 2022, and the wheat and rice markets in May 2022, which helps to understand the combined impact of the COVID-19 pandemic, frequent weather extremes, and the Russia-Ukraine conflict on the global food market more intuitively and deeply.

Based on the quantified downside and upside risk measures, we apply the K-S test to assess the significance of extreme risk spillover effects from agricultural futures to agricultural spots before and after the outbreak of the Russia-Ukraine conflict, as well as the possible asymmetries of these effects at the directional (downside versus upside) and temporal (pre-outbreak versus post-outbreak) levels. The test results suggest that for each grain crop, the futures market has significant downside and upside risk spillover effects on its corresponding spot market over both sub-periods, and there exists strength difference between these extreme risk spillover effects. Specifically, at the directional level, the soybean and maize markets consistently have stronger downside risk spillovers than upside risk spillovers over both periods. The downside risk spillovers in the wheat and rice markets are stronger than the upside risk spillovers before the outbreak of the conflict, but the opposite is true afterward. Furthermore, at the temporal level, we find that in the soybean market, both the downside and upside risk spillover effects over the pre-outbreak period are stronger than those over the post-outbreak period. For maize, there is basically no significant strength difference in the downside risk spillover effects between the two periods, but the upside risk spillover effect before the outbreak of the conflict is stronger than that afterward. In the wheat and rice markets, their downside risk spillover effects over the pre-outbreak period are stronger than those over the post-outbreak period, whereas their upside risk spillover effects over the post-outbreak period are stronger than those over the pre-outbreak period.

The combination of geopolitical turmoil, extreme weather, and supply chain bottlenecks caused by the COVID-19 pandemic has markedly increased uncertainty in the world economic recovery process. The subsequent outbreak and escalation of the Russia-Ukraine conflict have further brought new negative impacts to the global economy, aggravated the systemic risk of the world food system, and led to a steady deterioration of the food security situation. In fact, ensuring global food security is a complex and systematic project. Relevant international organizations, such as the WTO, FAO, and WFP, should work together to maintain the smooth operation of the international food market, advocate for food trade liberalization, and promote joint actions to ensure the unhindered and convenient circulation of food. Furthermore, countries and regions worldwide should enhance food governance and cooperation, refrain from food trade protectionism, and strive for practical solutions to address the excess volatility in food prices, aiming to establish a fair, open, and efficient global food supply system.

Agricultural futures are widely recognized as essential tools for producers, consumers, and investors to hedge price risks and diversify asset allocations. However, in recent years, amid the turbulent international situation and growing financialization of commodity markets, excessive speculation in agricultural futures markets has become prevalent, resulting in frequent and substantial price fluctuations. As our study findings indicate, risks in international agricultural futures markets have intensified and significantly spilled over to agricultural spot markets, amplifying systemic risk in the global food market and posing severe challenges to global food security. Using derivatives such as futures and options to hedge risks, guarantee farmers' income, and stabilize food markets is a global issue of common concern that has been explored in several countries worldwide, including ``crop insurance'' in the United States, ``insurance \& futures'' in China, ``producer put options'' in Brazil, and ``agricultural option subsidies'' in Mexico.

Therefore, countries and regions worldwide should strengthen the supervision and management of agricultural futures markets, jointly curb excessive speculation in food markets, and give full play to the basic functions of price discovery and risk diversification of agricultural futures, so as to provide a strong guarantee for international and domestic food security, especially during major events such as the Russia-Ukraine conflict. Additionally, futures exchanges are expected to continuously promote financial innovation, launch more varieties of agricultural futures and options, and explore the potential functions of market tools in safeguarding food security, further shouldering financial responsibility for food security. Moreover, considering the tail dependence and extreme risk spillovers among the agricultural futures-spot markets, investors are encouraged to appropriately add other categories of commodity futures when investing in agricultural futures to realize portfolio diversification and reduce investment risks.

It is recognized that models and their modes of application can always be improved. Both the selected single copula models and the self-constructed mixed copula models in our paper are static, and do not take dynamic tail dependence into consideration. In future research, time-varying copula models are expected to be introduced into our analytical framework, which may contribute to a more accurate and comprehensive description of tail dependence. Furthermore, the main focus of our work is the impact of the Russia-Ukraine conflict on the tail dependence structure and extreme risk spillover across pairwise futures and spot markets of a specific agricultural crop staple, without providing insights into tail dependence and risk spillover across different agricultural products. Bringing a cross-variety perspective to the current research could be regarded as a promising issue for extension. In conclusion, the impact of the Russia-Ukraine conflict on the global food market remains an under-studied topic that could benefit from further investigation. 

\section*{Acknowledgment}

 This work was supported by the National Natural Science Foundation of China (Grant Numbers: 72201099, 72171083), the Shanghai Outstanding Academic Leaders Plan, the Fundamental Research Funds for the Central Universities, and the China Postdoctoral Science Foundation (Grant Number: 2023T160217).

\section*{Data availability}

The data sets investigated in this paper are sourced from the Wind database (\url{https://www.wind.com.cn}) and the website of the International Grains Council (\url{https://www.igc.int}).

%

\end{document}